\documentclass[twocolumn,usenames,dvipsnames]{aastex63}

\accepted{\today}
\submitjournal{ApJ}

\usepackage{comment, makecell,array,multirow, chngcntr}

\usepackage{xcolor}

\def\la{\mathrel{\hbox{\rlap{\hbox{\lower4pt\hbox{$\sim$}}}\hbox{$<$}}}}
\def\ga{\mathrel{\hbox{\rlap{\hbox{\lower4pt\hbox{$\sim$}}}\hbox{$>$}}}}
\def\mic{$\mu$m}

\def\kms{km~s$^{-1}$}

\def\Afe{$A_{\rm Fe}$}
\def\nh{$N_{\rm H}$}

\def\dm15{{$\Delta$}$m_{15}$}

\def\v10{$V_{10}$(Si~II)}

\def\W575{$W(5750)$}
\def\W610{$W(6100)$}
\def\6100{the 6100~\AA\ absorption}

\def\m{M$_\odot$}
\def\msun{M$_\odot$}
\def\msunyr{M$_\odot$~y$^{-1}$}

\def\Ha{H$\alpha$\,}
\def\Hb{H$\beta$\,}

\def\ergs{erg~s$^{-1}$}
\def\gcm3{g~cm$^{-3}$}
\def\cm3{cm$^{-3}$}
\def\cm2g{cm$^{2}$~g$^{-1}$}

\def\degree{$^{\rm o}$}
\def\CaII7291{[Ca~{\sc II}] $\lambda\lambda$7291,7323}

\def\HeI{He~{\sc I} }

\def\OI6300{[O~{\sc I}] $\lambda\lambda$6300,6364}

\def \lta {\mathrel{\vcenter
     {\hbox{$<$}\nointerlineskip\hbox{$\sim$}}}}

\def\apj{ApJ}
\def\apjl{ApJL}
\def\apjs{ApJS}
\def\aj{AJ}

\def\mnras{MNRAS}
\def\aap{A\&A}
\def\pasp{PASP}

\def\araa{ARAA}

\def\swift{{\it Swift}}
\def\nustar{{\it NuSTAR}}
\def\chandra{{\it Chandra}}

\def\cxonustar{{\it CXO-NuSTAR}}

\def\vapec{{\tt vapec}}
\def\tbabs{{\tt tbabs}}

\def\cflux{{\tt cflux}}
\def\xspec{{\sc xspec}}
\def\xrtpipeline{{\sc xrtpipeline}}
\def\pimms{{\sc pimms}}
\def\ciao{{\sc ciao}}
\def\caldb{{\sc caldb}}
\def\xselect{{\sc xselect}}
\def\combine{{\sc combine\_spectra}}

\def\specextract{{\sc specextract}}
\def\nupipeline{{\sc nupipeline}}
\def\nuproducts{{\sc nuproducts}}

\newcommand{\sn}{SN~2014C}

\newcommand{\foiii}[1]{{{[\ion{O}{3}]} #1}}

\newcommand{\foi}[1]{{{[\ion{O}{1}]} #1}}

\newcommand{\fsii}[1]{{{[\ion{S}{2}]} #1}}
\newcommand{\fnii}[1]{{{[\ion{N}{2}]} #1}}
\newcommand{\fcaii}[1]{{{[\ion{Ca}{2}]} #1}}
\newcommand{\fneiii}[1]{{{[\ion{Ne}{3}]} #1}}

\begin{document}

\author[0000-0002-0977-1974]{Benjamin P. Thomas}
\affiliation{Department of Astronomy, University of Texas at Austin, 2515 Speedway, Stop C1400
Austin, Texas 78712-1205, USA}

\author[0000-0003-1349-6538]{J.\ Craig Wheeler}
\affiliation{Department of Astronomy, University of Texas at Austin, Austin, Texas}

\author[0000-0002-4661-7001]{Vikram V. Dwarkadas}
\affiliation{Department of Astronomy and Astrophysics, University of Chicago, 5640 S Ellis Ave, Chicago, Illinois, 60637}

\author{Christopher Stockdale}
\affiliation{Physics Department, Marquette University, Milwaukee, Wisconsin}

\author[0000-0001-8764-7832]{Jozsef Vink{\'o}}
\affiliation{ Konkoly Observatory,  CSFK, Konkoly-Thege M. \'ut 15-17, 
Budapest, 1121, Hungary}
\affiliation{ELTE E\"otv\"os Lor\'and University, Institute of Physics, P\'azm\'any P\'eter s\'et\'any 1/A, Budapest, 1117 Hungary}
\affiliation{Department of Optics \& Quantum Electronics, University of Szeged, D\'om t\'er 9, Szeged, 6720, Hungary}
\affiliation{Department of Astronomy, University of Texas at Austin, 2515 Speedway, Stop C1400
Austin, Texas 78712-1205, USA}

\author[0000-0003-4897-7833]{David Pooley}
\affiliation{Department of Physics and Astronomy, Trinity University, San Antonio, Texas}
\affiliation{Eureka Scientific, Inc.}

\author[0000-0002-2523-5485]{Yerong Xu}
\affiliation{Department of Astronomy and Astrophysics, University of Chicago, Chicago, Illinois}
\affil{Universit\`a degli Studi di Palermo, Dipartimento di Fisica e Chimica, via Archirafi 36, I-90123 Palermo, Italy}
\affil{INAF - IASF Palermo, Via U. La Malfa 153, I-90146 Palermo, Italy}

\author[0000-0003-2307-0629]{Greg Zeimann}
\affiliation{McDonald Observatory, University of Texas at Austin, Austin, Texas}

\author{Phillip MacQueen}
\affiliation{McDonald Observatory, University of Texas at Austin, Austin, Texas}

\shorttitle{SN 2014C}
\shortauthors{Thomas et al.}

\correspondingauthor{Benjamin P. Thomas}
\email{bpthomas@utexas.edu}

\graphicspath{{./}{}}

\title{Seven Years of SN 2014C: a Multi-Wavelength Synthesis of an Extraordinary Supernova}

\begin{abstract}

SN~2014C was originally classified as a Type Ib supernova, but at phase $\phi = 127$~d post-explosion strong \Ha\ emission was observed. SN~2014C has since been observed in radio, infrared, optical and X-ray bands. Here we present new optical spectroscopic and photometric data spanning $\phi = 947 - 2494$~d post-explosion. We address the evolution of the broadened \Ha\ emission line, as well as broad [\ion{O}{3}] emission and other lines. We also conduct a parallel analysis of all publicly available multi-wavelength data.  From our spectra, we find a nearly constant \Ha\ FWHM velocity width of ${\sim}2000$ \kms\ that is significantly lower than that of other broadened atomic transitions (${\sim}3000-7000$ \kms) present in our spectra ([\ion{O}{1}] $\lambda 6300$; [\ion{O}{3}] $\lambda \lambda 4959, 5007$; \ion{He}{1} $\lambda 7065$; [\ion{Ca}{2}] $\lambda\lambda 7291, 7324$). The late radio data demand a fast forward shock (${\sim}10,000$ km s$^{-1}$ at $\phi = 1700$~d) in rarified matter that contrasts with the modest velocity of the \Ha. We propose that the infrared flux originates from a toroidal-like structure of hydrogen surrounding the progenitor system, while later emission at other wavelengths (radio, X-ray) likely originates predominantly from the reverse shock in the ejecta and the forward shock in the quasi-spherical progenitor He wind. We propose that the \Ha\ emission arises in the boundary layer between the ejecta and torus. We also consider the possible roles of a pulsar and a binary companion.

\end{abstract}

\keywords{supernovae: general - supernovae: individual (SN~2014C) - circumstellar matter - X-rays: individual (SN~2014C) - radio continuum: general 
}

\section{Introduction} \label{sec:intro}

SN~2014C was discovered in the nearby ($d_{\rm L} = 14.7 \pm 0.6$ Mpc; Freedman et al. 2001) spiral galaxy NGC 7331 on 5 January 2014 by the Lick Observatory Supernova Search \citep{Kim14}. \citet{Margutti17} estimate the time of first light to be 30 December 2013. Maximum {\it V}-band magnitude was reached on 13 January 2014 \citep{Milisavljevic15}. 
SN~2014C was first observed \citep{Milisavljevic15} as a photometrically and spectroscopically normal stripped-envelope supernova \citep{clocc97} of Type Ib that showed little photospheric evidence for hydrogen, but substantial evidence for helium. Solar occlusion imposed an observational hiatus, but 127 days after first light, SN~2014C showed a prominent emission line of \Ha\ suggesting that some time prior to that the ejecta had collided with a hydrogen-rich circumstellar medium (CSM). SN~2014C also became a prominent source of X-ray, radio, and infrared emission \citep{tinyan16,Margutti17,Anderson17,Bietenholz18,tinyan19, Brethauer20,Bietenholz21}

The transformation of SN~2014C from a Type~Ib to revealing evidence for collision with hydrogen-rich material is consistent with a helium star that exploded in a relatively low-density cavity and then collided with matter representing the previously-ejected envelope of the progenitor. The ejecta continued to interact with the CSM for at least 5 years after the explosion \citep{tinyan19}.

The 15.7 GHz radio light curve reported by \citet{Anderson17} and Spitzer IR observations \citep{tinyan16, tinyan19} provided the only data on SN~2014C during the first solar occlusion. The radio data showed a first peak at 80 d after first light. \citet{Anderson17} estimated the second radio rise to start at 186 d. The second peak reached maximum about 400 days after first light.

\citet{Bietenholz18} employed Very Long Baseline Interferometry (VLBI) to measure the rate of change of the size of an annulus of emission detected on their radio images and the associated velocity of the shock front. The first epoch at 384 d after first light indicated a substantial slowing compared to the photospheric velocity of the supernova \citep{Margutti17}, presumably due to interaction with the CSM. A second epoch at 1057 d suggested a constant rate of expansion between the two epochs of ${\sim}13,600$ \kms. The image in the second epoch was essentially round, but marked by a bright spot in the West. To within uncertainties of ${\sim}2400$~\kms\ the centroid showed no proper motion. \citet{Bietenholz21} found that the VLBI image at 5 years after the explosion was consistent with a spherical shell.

\citet{tinyan19} examined the conditions in SN~2014C with infrared photometry and spectroscopy from one to five years past the explosion. They found intermediate-width \HeI 1.083 \mic\ emission from the interaction region up to 1639 days post-explosion and confirmed ongoing CSI at 1920 days with {\it Spitzer} photometry. They assumed that the IR light curve was representative of the bolometric light curve. They argued that the light curve after 500 days is consistent with a model in which the supernova collides with a CSM produced by a wind of constant velocity and mass loss rate of ${\sim}10^{-3}$ \msunyr\ that represents an additional CSM component exterior to the high-density shell invoked by \citet{Milisavljevic15} and \citet{Margutti17}.

\citet{harris20} used one-dimensional hydrodynamic models of supernova ejecta colliding with a dense shell to explore the nature of SN~2014C. They found that shells of substantial density contrast can lead to departures from self-similar behaviour. They note that ejecta can be slowed significantly by a relatively dense shell even if it has rather small mass and hence that low line velocities do not necessarily represent massive shells. They point out that the radio rise at about 186 days is significantly after the first detected \Ha\ emission at 127 days and propose that the early rise in radio flux occurred after the forward shock had departed the proposed dense shell and was propagating in the outer CSM. They derive a significantly smaller mass of the dense shell, ${\sim}0.05$ \msun, than do \citet{Margutti17}.

\citet{sunmc20} present HST observations of the star cluster that hosted SN~2014C. From the spectral energy distribution, they derive a cluster age of 20 Myr. If the progenitor star of SN~2014C was coeval with the cluster, it would have a mass of about 12 \msun. \citet{sunmc20} argue that if the progenitor were a single star of this mass, it would not have ejected its hydrogen envelope and thus could not have exploded as a SN~Ib. They construct binary evolution models for which the progenitor could have had a ZAMS mass of 11 \msun\ and lost its envelope in Case B/C or Case C mass transfer \citep{kipp90}. The common envelope mass loss rate is ${\sim}10^{-3}$ \msunyr, comparable to that deduced by \citet{tinyan19}. From the bolometric light curve and the diffusion theory of \citet{arnett82}, the ejected mass would be ${\sim}2$\msun. With the addition of a neutron star of mass ${\sim}1.4$\msun, the total mass of the helium star progenitor would be ${\sim}3$\msun, consistent with the estimates of \citet{Milisavljevic15} and \citet{Margutti17}. \citet{sunmc20} note a caveat to this conclusion if the estimates of the opacity associated with the light curve are too low due to helium and oxygen recombination \citep{wheelerjc15,maund18,kk19}. 

The focus of this paper is a presentation and interpretation of seven years of optical IFU spectroscopy and imaging and narrow-band imaging of \sn\ and its immediate environment. We present a detailed discussion of the reduction process for the IFU data. We also do our own reduction and analysis of all the available X-ray data and present a further epoch of radio data at 2063 days after first light. To obtain a complete representation of the data on \sn, we gathered and analyzed publicly-available optical and IR spectroscopy and radio data. 

While some of these data have been available for years, our optical data is new, calling for a complete synthesis of all available data. Our study of these data reveal that a comprehensive integration of all the multi-epoch and multi-wavelength data required a departure from spherical symmetry. A particular conundrum that emerged is the discrepant behavior of the detected \Ha\ emission that revealed only a slight decrease of luminosity with time compared with other bands and a lower velocity than other emission lines and the velocity implied by VLBI imaging. We propose that these discrepancies can be resolved with a multi-component, non-spherical configuration of the environment of \sn.

The structure of this paper is as follows: \S \ref{sec:obs} presents our optical imaging and spectroscopic observations and an analysis of emission line profiles; \S \ref{sec:xray} gives our analysis of public X-ray data; \S \ref{sec:radio} presents our recent VLA observation; \S \ref{sec:syn} synthesizes the multiwavelength data from \sn\ and presents the argument for and analysis of a scenario in which both a dense CSM torus and a low density quasi-spherical wind distribution are required to account for the observations; \S \ref{sec:concl} summarizes our conclusions. Throughout this work we assume a flat $\Lambda$CDM cosmology with $\Omega_{\rm M} = 0.3$ and $H_0 = 71\, {\rm km\, s}^{-1}{\rm Mpc}^{-1}$. Observations that are new with this paper are summarised in Table \ref{tab:obs_summary}.

\begin{deluxetable*}{rrrrrrr}
\label{tab:obs_summary}
\tablecaption{Summary of all observations used in this work. X-ray data were re-reduced and analysed for this work.
}
\tablehead{ & \colhead{Telescope/inst.} & \colhead{Filter} & \colhead{Nobs} & \colhead{Date} & \colhead{Phase} & \colhead{Reference} }
\startdata
\hline
X-ray & NuStar & 3-79 keV & 7 & 2015-08-29 to 2020-04-30 & 607 to 2306 & This work (PI: Margutti) \\
      & CXO & 0.3-10 keV &  9 & 2015-08-28 to 2020-04-18 & 606 to 2294 & This work (PI: Margutti) \\
      & NuStar & 3-79 keV & 2 & 2015-01-29 to 2015-04-19 & 394 to 475 & \cite{Margutti17} \\
      & CXO & 0.3-10 keV &  3 & 2014-11-03 to 2015-04-20 & 307 to 476 &  \cite{Margutti17} \\
      & Swift & 0.2-10 keV & 1 & 2014-01-06 to 2014-01-19 & 7 to 10 & \cite{Margutti17} \\
\hline
Optical & HET/LRS2 & Spec. & 8 & 2016-08-06 to 2020-11-05 & 947 to 2494 & This work (PI: Wheeler) \\
        & HJS/DIAFI & \Ha-narrow & 5 & 2015-05-15 to 2018-08-31 & 500 to 1700 & This work (PI: Wheeler) \\
        & LBT/MODS & Spec. & 1 & 2014-10-22 & 295 & \cite{Milisavljevic15} \\
        & Keck/DEIMOS & Spec. & 3 & 2014-10-02 to 2017-08-18 & 275 to 1323 & \cite{mauerhan18} \\
        & Keck/LRIS & Spec. & 4 & 2014-07-29 to 2015-09-16 & 210 to 623 & \cite{mauerhan18} \\
        & MMT/Blue Ch. & Spec. & 3 & 2014-05-06 to 2015-04-25 & 127 to 479 & \cite{Milisavljevic15} \\
        & Lick/Kast & Spec. & 6 & 2014-01-22 to 2014-08-28 & 23 to 240 & \cite{mauerhan18} \\
        & FLWO/FAST & Spec. & 1 & 2014-01-09 & 10 & \cite{Milisavljevic15} \\
\hline  
IR  & Keck/NIRES & Spec. & 1 & 2018-09-02 & 1702 & \cite{tinyan19} \\
    & Gemini/NIRI  & L' & 1 & 2018-06-18 & 1626 & \cite{tinyan19} \\
    & Gemini/NIRI  & M' & 1 & 2018-06-18 & 1626 & \cite{tinyan19} \\
    & P200/WIRC & J & 3 & 2017-09-30 to 2018-07-17 & 1366 to 1655 & \cite{tinyan19} \\
    & Keck/MOSFIRE & Spec. & 1 & 2017-09-28 & 1364 & \cite{tinyan19} \\
    & P200/TripleSpec & Spec. & 2 & 2017-08-09 to 2018-06-22 & 1314 to 1630 & \cite{tinyan19} \\
    & P200/WIRC & H & 4 & 2017-07-10 to 2018-07-27 & 1284 to 1665 & \cite{tinyan19} \\
    & NOT/NOTCam & H & 1 & 2015-09-25 & 632 & \cite{tinyan19} \\
    & NOT/NOTCam & J & 1 & 2015-09-25 & 632 & \cite{tinyan19} \\
    & NOT/NOTCam & Ks & 1 & 2015-09-25 & 632 & \cite{tinyan19} \\
    & P200/WIRC & Ks & 5 & 2014-10-20 to 2018-07-17 & 294 to 1655 & \cite{tinyan19} \\
    & Spitzer/IRAC & 3.6\mic\ & 16 & 2014-02-21 to 2019-04-08 & 53 to 1919 & \cite{tinyan19} \\
    & Spitzer/IRAC & 4.5\mic\ & 16 & 2014-02-21 to 2019-04-11 & 53 to 1922 & \cite{tinyan19} \\
\hline
Radio & VLA & 15.1 GHz & 1 & 2020-05-06 & 2323 & \cite{Bietenholz21} \\
      & VLA & 9 GHz & 1 & 2019-08-31 & 2063 & This work (PI: Stockdale) \\
      & eMerlin & 1.5 GHz & 1 & 2015-05-04 & 489 & \cite{Anderson17} \\
      & eMerlin & 5.1 GHz & 2 & 2015-04-18 to 2015-05-06 & 473 to 491 & \cite{Anderson17} \\
      & VLBA & 8.4 GHz & 4 & 2015-01-17 to 2016-10-20 & 384 to 1057 & \cite{Bietenholz18} \\
      & VLBA & 22.1 GHz & 1 & 2015-01-17 & 384 & \cite{Bietenholz18} \\
      & eMerlin & 5.5 GHz & 1 & 2014-01-19 & 20 & \cite{Anderson17} \\
      & AMI & 15.7 GHz & 81 & 2014-01-15 to 2015-07-20 & 17 to 567 & \cite{Anderson17} \\
      & JVLA & 4.9 GHz & 12 & 2014-01-11 to 2020-04-02 & 12 to 2278 & \cite{Bietenholz21} \\
      & JVLA & 7.1 GHz & 13 & 2014-01-11 to 2020-04-24 & 12 to 2300 & \cite{Bietenholz21} \\
\hline                                                                                                                         
    \enddata
\end{deluxetable*}

\section{Optical Observations} \label{sec:obs}

\begin{deluxetable*}{rcrrrc}
\label{tab:record}
\tablewidth{0pt}
\tablecolumns{6}
\tablecaption{ New data of SN 2014C }
\tablehead{ \colhead{Date} & \colhead{$\phi$} & \colhead{Telescope} & \colhead{Bandpass} & \colhead{Luminosity} & \colhead{Exposure time} \\
\vspace{-.7cm} \\
\colhead{} & \colhead{(rest-frame days)} &  \colhead{/Instrument} & \colhead{} & \colhead{($10^{38}$ \ergs)} & \colhead{(s)}}
\startdata
    2015-05-20 & 505 & HJS/DIAFI & \Ha-narrow & $8.47 \pm 1.79$ & 150 \\
    2015-08-23 & 599 & HJS/DIAFI & \Ha-narrow & $7.23 \pm 1.87$ & 300 \\
    2016-06-07 & 888 & HJS/DIAFI & \Ha-narrow & $6.72 \pm 1.97$ & 142 \\
    2016-08-06 & 947 & HET/LRS2-R & \tablenotemark{a} & $7.79^{+0.71}_{-0.52}$ & 1800 \\
    2016-09-04 & 976 & HET/LRS2-B & \tablenotemark{a} &  $6.64^{+0.25}_{-0.15}$ & 1800 \\
    2016-09-24 & 996 & HJS/DIAFI & \Ha-narrow & $5.16 \pm 1.70$ & 150 \\
    2017-05-24 & 1237 & HET/LRS2-R & \tablenotemark{a} & $8.74^{+0.57}_{-0.51}$ & 1800 \\
    2017-08-17 & 1322 & HET/LRS2-B & \tablenotemark{a} & $7.28^{+0.31}_{-0.21}$ & 2000 \\
    2018-06-11 & 1619 & HET/LRS2-B & \tablenotemark{a} & $5.73^{+3.70}_{-1.49}$ & 1800 \\
    2018-09-05 & 1705 & HJS/DIAFI & \Ha-narrow & $6.39 \pm 1.77$ & 142 \\
    2019-08-25 & 2057 & HET/LRS2-B & \tablenotemark{a} & $5.49^{+0.39}_{-0.20}$ & 2200\\
    2019-08-31 & 2063 & VLA & 9 GHz & 0.287 & 2380 \\
    2020-05-30 & 2336 & HET/LRS2-B & \tablenotemark{a} & $4.34^{+1.23}_{-0.26}$ & 1800 \\
    2020-11-04 & 2493 & HET/LRS2-B & \tablenotemark{a} & $5.23^{+2.13}_{-0.90}$ & 3600 \\
    2020-11-05 & 2494 & HET/LRS2-R & \tablenotemark{a} & $5.12^{+0.23}_{-0.20}$ & 3600 \\ 
\enddata
\tablenotetext{a}{Luminosity was derived via Gaussian decompositions to the broadened \Ha\ spectral profile. These correspond to luminosities of the broadened \Ha\ component only.} 

\end{deluxetable*}

\subsection{DIAFI images}
 \label{subsec:diafi_images}
We utilize the Direct Imaging Auxiliary Functions Instrument (DIAFI\footnote{https://mcdonald.utexas.edu/for-researchers/research-facilities/2-7-m-107-harlan-j-smith-telescope/165-researchers/643-diafi}) imager on the Harlan J. Smith 2.7 m telescope at McDonald Observatory since 2014 February to search for supernovae exhibiting evidence of delayed collision and excitation of \Ha\ with narrowband filters, one near the expected redshifted wavelength of \Ha\ ($\lambda_{\rm central} = 6585 {\rm \AA}$, FWHM = 70 ${\rm \AA}$) and another in an ``off” band ($\lambda_{\rm central} = 6675 {\rm \AA}$, FWHM = 70 ${\rm \AA}$) for calibration. Procedures for reducing the DIAFI data are presented in \S 3 of \citet{Vinko17} that also summarized our earliest results. Among other results, we confirmed the broad \Ha\ in SN~2014C previously reported by \citet{Milisavljevic15}.

\begin{figure}
    \centering
    \includegraphics[width=\columnwidth]{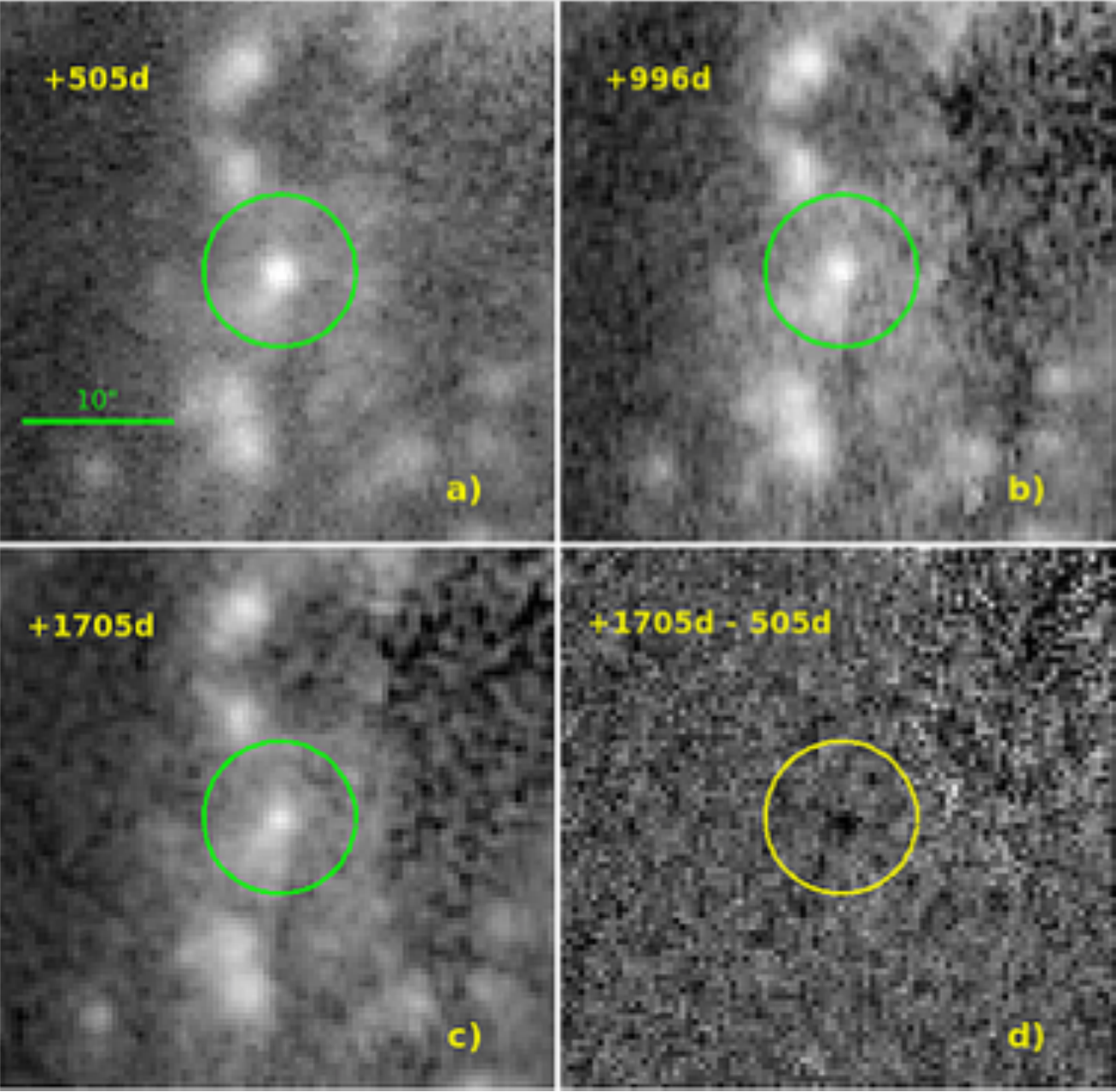}
    \caption{{\it Panels a, b,} and {\it c}: Narrow-band, continuum-subtracted \Ha\ images of the field of SN~2014C taken with the DIAFI camera. The center of the host galaxy, NGC~7331, is located to the upper right, slightly off the illustrated frames. The phase of each observation is shown in the top left corner. The location of SN~2014C is given by the green circle. The supernova falls within an extended H II region and appears as a point source. The fading of the emission peak with respect to the flux of nearby H II regions is apparent.
    {\it Panel d}: The difference image of the frames shown in Panel c and Panel a. The dark spot at the supernova position indicates reduced \Ha\ flux from \sn\ on the $\phi = 1705$ d frame with respect to the $\phi = 505$ d frame. These observations show that between 2015 and 2018 the \Ha\ line flux from \sn\ decreased substantially. 
    }
    \label{fig:image}
\end{figure}

SN~2014C exploded in a spiral arm of NGC~7331 that is rich in H II regions. We define a temporal phase parameter, $\phi$, taken to be rest-frame days from first light (2013-12-30) as determined by \citet{Margutti17}, and refer to all data with $\phi$.
Figure \ref{fig:image} shows a narrow-band image of the field of SN~2014C taken with the DIAFI camera at three epochs, $\phi =$ 305, 996, and 1705 d, illustrating the fading of the supernova. By that third epoch, the supernova had clearly faded but still showed spectral evidence for a broadened component of \Ha\ (see \S \ref{subsec:spectra}).  The location of the supernova, shown in the green circle in each panel, falls within one of the ambient H II regions. The image of that H II region in Panel c is clearly extended rather than point-like. Panel d of Figure \ref{fig:image} shows a subtraction of the image in Panel a obtained in 2015 at $\phi = 505$ d from the image in Panel c obtained in 2018 at $\phi = 1705$ d. The majority of the field subtracts very cleanly. 
All the images of the H II regions are gone, including that within which SN~2014C exploded. The image of the supernova in Panel d shows as a well-resolved dark point, establishing that 
the \Ha\ flux coming from the vicinity of the supernova was substantially less in 2018 than in 2015. The decrease of the \Ha\ flux from the site of \sn\ suggests that the source of the \Ha\ photons is still the ejecta-CSM interaction, but the \Ha\ excitation process has substantially decreased since the start of the interaction.

\subsection{HET/LRS2 IFU Reduction}
\label{sec:lrs2reduction}

\begin{figure*}
    \centering
    \includegraphics[width=\textwidth]{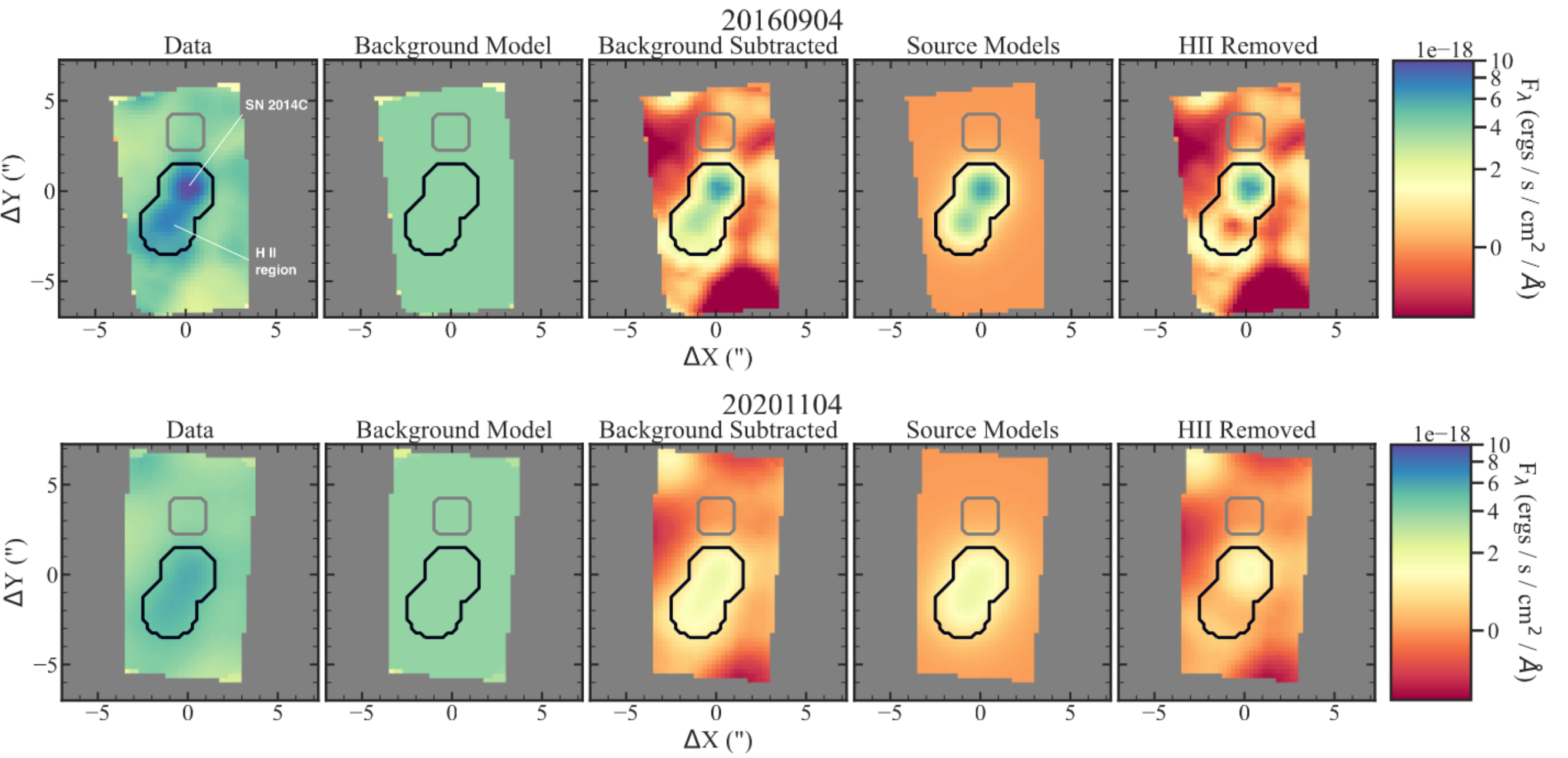}
    \caption{Diagnostic LRS2 IFU images for data on SN~2014C from $\phi = 976$ d (top) and $\phi = 2493$ d (bottom). The scale is given in arcseconds. The images are centered on 6583 \AA\ and collapsed over a 20 \AA\ window using a Gaussian-weighted average ($\sigma = 6$ \AA).  The first panel shows the total data from the region revealing both SN~2014C as the central object and a nearby (${\sim}150$ pc distant) H II knot to the lower left of the supernova that is also revealed in Figure \ref{fig:image}. The SN emission and the nearby H II region are indicated on the first panel. The second panel represents the data from the background captured in the small square near the top of the image. The third panel presents the data from which the background is subtracted. The fourth panel gives the source models for SN~2014C and the spatially-resolved H II region. The fifth panel shows the original data corrected for the background and with the H II region removed. 
    }
    \label{fig:IFUimage}
\end{figure*}

The LRS2 IFU image extraction process enables another means to image the environment of SN~2014C (in addition to our DIAFI imaging described in \S \ref{subsec:diafi_images}). Figure \ref{fig:IFUimage} shows the data from the supernova and a nearby H II region. This image can be compared to Figure 2 of \citet{Milisavljevic15}.

 The spectra and IFU images of SN~2014C reported here were obtained with the Low-Resolution Spectrograph 2 (LRS2; \citealt{Chonis16}) on the 10m Hobby–Eberly Telescope (HET; \citealt{het}, \citealt{het2}). LRS2 comprises two IFU spectrographs separated by 100 arcseconds on sky: LRS2-B (3650\AA\ - 6950\AA) and LRS2-R (6450\AA\ - 10500\AA).  Each spectrograph has 280 fibers covering  6"×12" with unity fill factor \citep{Chonis16}.  We use the HET LRS2 pipeline, Panacea\footnote{https://github.com/grzeimann/Panacea}, to perform the initial reductions including fiber extraction, wavelength calibration, astrometry, and flux calibration. This reduction and calibration process is visualized in Figure \ref{fig:IFUimage}. There are two channels for each spectrograph: UV and Orange for LRS2-B and Red and Farred for LRS2-R.  Before November 2016, the UV channel had to be zeroed out due to a failed UV chip that was replaced on this date.  On each exposure, we combine fiber spectra from the two channels into a single data cube accounting for differential atmospheric refraction.  We then identify the target \sn\ in each observation and rectify the data cubes to a common sky coordinate grid with \sn\ at the center.

Since the IFU contains background light from the host galaxy we use separate ``blank'' observations for sky subtraction scaled to the relevant exposure.  We examine the residuals manually near bright sky lines to calculate this scalar factor.  We then subtract the scaled sky from each data cube.  At this stage, the data cubes still include light from the background galaxy, which we can use to improve our initial flux calibration.  We define a common region in each data cube (grey box from Figure \ref{fig:IFUimage}) far enough from SN~2014C that it should be unaffected by our target. In each of our observations, the median spectrum in this region should be unchanging thus allowing us to use it as a common scale for flux calibration.  We measure the median spectrum in the grey box from Figure \ref{fig:IFUimage} and scale that spectrum to the average spectrum from all observations.  We restrict the normalization calculation to wavelengths in common to both LRS2-B and LRS2-R (6450\AA\ - 6850\AA).  We then apply this normalization to our data cubes.  The normalization factors were typically between 0.9 - 1.1.  
\sn\ is near an H II knot within the larger H II complex surrounding the supernova.  The H II knot is separated from our target by 2.15", and the seeing conditions across the observations range from 1.6"-3.0".  We chose to model the SN and the H II knot simultaneously and mask the two sources for host galaxy background subtraction.  We use a Gaussian kernel with a $\sigma$=1.75" to spatially smooth and interpolate the background light over our masked sources.  We then subtracted our smoothed background model.

For each observation, we simultaneously model the \sn\ source and the H II knot with a Moffat profile in an image collapsed about observed \Ha. The Moffat profiles had FWHM that ranged from 1.6-3.0".  We then fix the Moffat models leaving only the amplitude of the two profiles free. At each wavelength of our data cubes, we fit the two free amplitudes to create 3-D models of \sn\ and the H II knot.  We use the 3-D model of the H II knot for two purposes: the sum of the model at each wavelength provides the spectrum for the H II region and we use the model to subtract the knot from the IFU observation.  After we subtract the H II region model from our data cube, we then use a 1.5" radius aperture for the spectral extraction of \sn.  We extrapolate the aperture spectrum to a total flux spectrum using our Moffat model.      

The normalization correction to go from the 1.5" aperture extraction to a total flux is the dominant uncertainty in the flux calibration. Taking the uncertainty in the normalization correction into account gives a rough measure of the \Ha\ flux as a function of time. The distribution of the normalization corrections is not Gaussian but can be characterized by the values exceeding a given percentile of the distribution with the 50th percentile representing the median of the distribution. The resulting percentile values of the correction distribution for each of our LRS2 observations of \sn\ are given in Table \ref{tab:normcorr_ha5}.

\begin{deluxetable}{r|rrr}
\label{tab:normcorr_ha5}
\tablecaption{Normalization correction for \Ha\ lines as a function of epoch for our LRS2 HET data. The corrections are given at the 50th percentile, the 16th, and the 84th. 
}
\tablehead{\colhead{epoch} & \colhead{correction} & \colhead{correction} & \colhead{correction} \\
\colhead{}  & 
\colhead{(50th)} & 
\colhead{(16th)} & 
\colhead{(84th)}
}
    \startdata
2016-08-06 & 1.64 & 1.53 & 1.79 \\
2016-09-04 & 1.31 & 1.28 & 1.36 \\
2017-05-24 & 1.53 & 1.44 & 1.63 \\
2017-08-17 & 1.40 & 1.36 & 1.46 \\
2018-06-11 & 2.65 & 1.96 & 4.36 \\
2019-08-25 & 1.40 & 1.35 & 1.50 \\
2020-05-30 & 1.69 & 1.59 & 2.17 \\
2020-11-04 & 2.51 & 2.08 & 3.53 \\
2020-11-05 & 1.54 & 1.48 & 1.61 \\
    \enddata
\end{deluxetable}

\subsection{Spectra} \label{subsec:spectra}

 The average resolving power of our LRS2 spectrograph is R$\sim$1500. The spectral resolutions, deduced from the FWHM of narrow spectral lamp lines, are 5.09 \AA\ and 4.24 \AA\ for the orange arm of LRS2-B and the red arm of LRS2-R, respectively. These correspond to ${\sim}300$ and ${\sim}250$~km~s$^{-1}$ velocity uncertainties at 5000~\AA, while in the vicinity of \Ha\ they are ${\sim}230$ and ${\sim}195$~km~s$^{-1}$, respectively.

We acquired nine spectra of \sn\ with our HET/LRS2 IFU set-up from 2016-08-06 through to 2020-11-05 corresponding to phases 947 to 2493 days after first light. Other optical spectra have been presented by \citet{Milisavljevic15}, \citet{Anderson17}, and \citet{mauerhan18}. IR spectra were given by \citet{tinyan19}. Table \ref{tab:record} gives information about new data acquired in our program, including the conversion from observing date to the temporal phase parameter, $\phi$, taken to be rest-frame days from first light (2013-12-30) as determined by \citet{Margutti17}.

\begin{figure*}
    \centering
    
    \includegraphics[width=\textwidth]{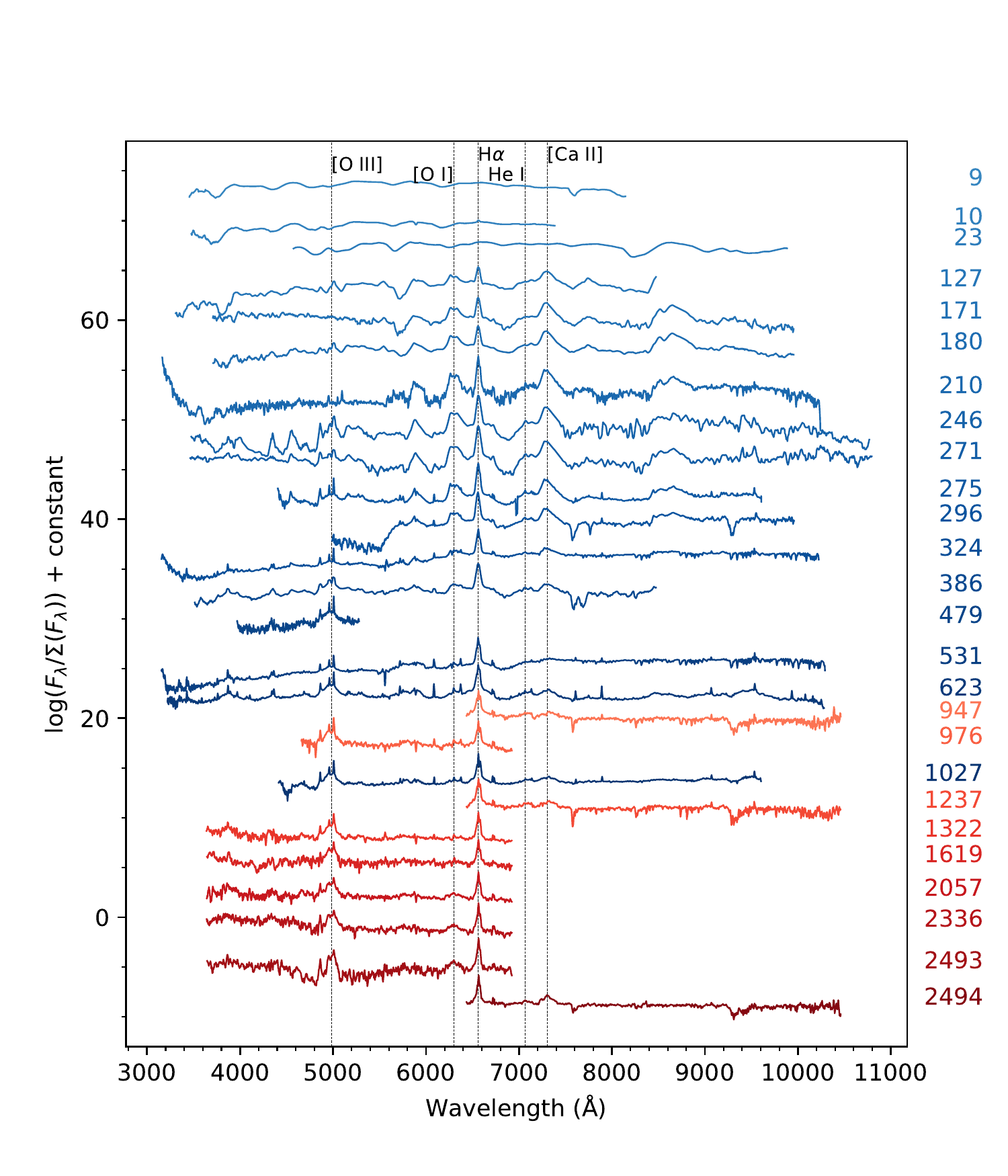}
    \caption{Twenty-six optical spectra of SN~2014C, including 17 publicly available spectra, and nine spectra obtained with our HET/LRS2 set-up from 2018-08-06 to 2020-11-05. The rest frame phase ($\phi$) from first light (2013-12-30, as determined by \citealt{Margutti17}) is provided on the right-hand side. Broadened emission lines that are pertinent to our analysis are identified with dashed vertical lines. Note the discernible broader components around 5000 \AA\ and around \Ha\ in the data after $\phi = 600$~d.}
    \label{fig:hetfull}
\end{figure*}

Figure \ref{fig:hetfull} presents the array of nine optical spectra of SN~2014C along with other optical data from the literature. The first ($\phi = 947$~d), third ($\phi = 1237$~d) and final ($\phi = 2494$~d) of our spectra were obtained with LRS2-R; the remainder were obtained with LRS2-B. Both instrumental components contain the \Ha/\fnii complex. 

The HET spectra at $\phi >947$~d reveal broad components to the \fcaii\ $\lambda \lambda$ 7291, 7324, \foi\ $\lambda$ 6300, \foiii\ $\lambda \lambda$ 4959, 5007 and \Ha\ emission. Evidence of broad emission from \fneiii\ $\lambda$ 3970 and H$\gamma$/\foiii\ $\lambda \lambda$ 4340, 4363 is also present, albeit at lower signal-to-noise ratio. 

Figure \ref{fig:hetfull} shows that standard nebular features of SN~Ib are visible in \sn. Among these are \foi\ $\lambda\lambda$6300, 6363; \fcaii\ $\lambda\lambda$7291, 7324; O I $\lambda$7774; and the Ca II IR triplet (Mg I] $\lambda$4571 is difficult to discern). These features that are produced in the inner ejecta are visible from $\phi = 127$~d to at least $\phi = 275$~d. Their presence means that the whole outer CSM is optically thin during that epoch, at least along the line of sight. We see none of these features in our data; they are basically gone by $\phi = 531$~d. The more highly-ionized \foiii\ appears after $\phi = 246$~d.

\subsubsection{Line Profiles} \label{subsubsec:profiles}

The core of our optical analysis lies in decomposing the blended and broadened emission line profiles into their various components. We assume that the individual components follow Gaussian distributions and combine these Gaussian distributions to compute a model emission complex. Each Gaussian is described by three free parameters that quantify the amplitude, mean and standard deviation. For example, for a quadruple Gaussian blend (that we use for both the \Ha\ and the \foiii/\Hb profiles) we have twelve free parameters, with an additional baseline parameter added to the full superposition for a total of thirteen free parameters.

To fit this model emission complex to the data, we use a Markov Chain Monte Carlo (MCMC) method implemented in the Python package emcee\footnote{\href{https://emcee.readthedocs.io/}{https://emcee.readthedocs.io/}}. For the \Ha\ complex, we use four components representing the broad \Ha, the narrow \Ha, and the two \fnii\ $\lambda \lambda$ 6548, 6583 lines. We initiate 30 walkers for 5000 steps and a burn-in period of 3000 steps. We use uniform prior distributions for all parameters with bounds informed by the observed data. For the \foiii lines, we also employ four components representing the broad and narrow components of the $\lambda \lambda$ 4959, 5007 \AA\ transitions. We use a similar method for \foi\ $\lambda$ 6300, \fcaii\ $\lambda\lambda 7291, 7324$ and He I $\lambda 10830$, where the latter IR spectra are presented in \citealt{tinyan19}. 

In addition to using Gaussian distributions to fit the \Ha\ broad component, we attempted to improve the fit with a Lorentzian distribution (while keeping the Gaussian for the three narrow components). We found that Lorentzian fits produced a comparable or worse $\chi^2$ per degree of freedom value relative to the corresponding Gaussian fits at all epochs. In reality, it is probable that there are contributions to the underlying profile broadening from both electron scattering and the velocity distribution of the emitting H atoms. Our aim is to measure the flux and FWHM of the various components to determine the luminosity and astrophysical source of that flux by comparing, for example, the \Ha\ to the \foiii emission. From hereon we adopt the Gaussian model as sufficiently representative of the broadened \Ha\ component.

An example of our \Ha\ decomposition at $\phi = 1322$~d is shown in Figure \ref{fig:benfit}. The full MCMC posterior distribution of all parameters from the same fit is given in Appendix \ref{sec:posterior}. These decompositions allow us to compute two critical quantities for our analysis: (1) the integrated flux (and hence luminosity) of each of the various components and (2) the FWHM of those components from which velocity information is conventionally derived.

We are primarily interested in the broadened \Ha\ relative to the other three components as it is most likely indicative of activity related to the supernova. We derive integrated fluxes and FWHM values of the broadened component from our quadruple Gaussian fits. We give the derived FWHM and corresponding velocity widths and respective  uncertainties in Table \ref{tab:fwhm_ha}.

\begin{deluxetable}{r|rrrr}
\label{tab:fwhm_ha}
\tablecaption{Derived full-width half maxima and the corresponding velocity widths of the broadened \Ha\ component from our HET/LRS2 spectra.
}
\tablehead{\colhead{$\phi$} & \colhead{FWHM} & \colhead{$\Delta$FWHM\tablenotemark{a}} &\colhead{$v$} & \colhead{$\Delta v$} \\
\colhead{(days)} & ($\rm{\AA}$) & ($\rm{\AA}$) & (\kms) & (\kms) 
}
\startdata
 947 & 51.9 & 4.2 & 2370 & 230 \\
 976 & 50.1 & 5.1 & 2290 & 260 \\
1237 & 50.4 & 4.2 & 2300 & 230 \\
1322 & 46.3 & 5.1 & 2120 & 260 \\
1619 & 44.8 & 5.1 & 2050 & 250 \\
2057 & 38.1 & 5.1 & 1740 & 250 \\
2336 & 46.2 & 5.1 & 2110 & 260 \\
2493 & 31.4 & 5.1 & 1440 & 240 \\
2494 & 34.7 & 4.2 & 1590 & 210 \\
\enddata
\tablenotetext{a}{Uncertainties quoted here are the quadrature sum of the error from the fit and the error from the spectral resolution.}
\end{deluxetable}

At $\phi=947$~d we find a broadened \Ha\ flux of $2.97\times10^{-14}$ erg s$^{-1}$ cm$^{-2}$ with a $7\%$ error from the flux calibration. There are several lines of evidence that the flux declines over the course of our observations. Although the uncertainties in the \Ha\ flux measured by the integrated flux in our spectra are relatively large, the flux measured in that way tends downward with time to within one or two sigma. That variation may not be statistically significant, but our DIAFI images (Figure \ref{fig:image}) provide an independently-derived line of evidence of that decline from a completely different technique that corrects for effects like seeing. 

At $\phi=947$~d we derive a line width value of FWHM = $51.9$ \AA\ with a $<2\%$ error from the fit (the error contribution from the spectral resolution can be as high as ${\sim}10\%$). The width of the broad \Ha\ component also remains effectively constant across all observed epochs with slight variability that may be attributed to the shot noise on the spectrum.

\begin{figure}
    \centering
    \includegraphics[width=\columnwidth]{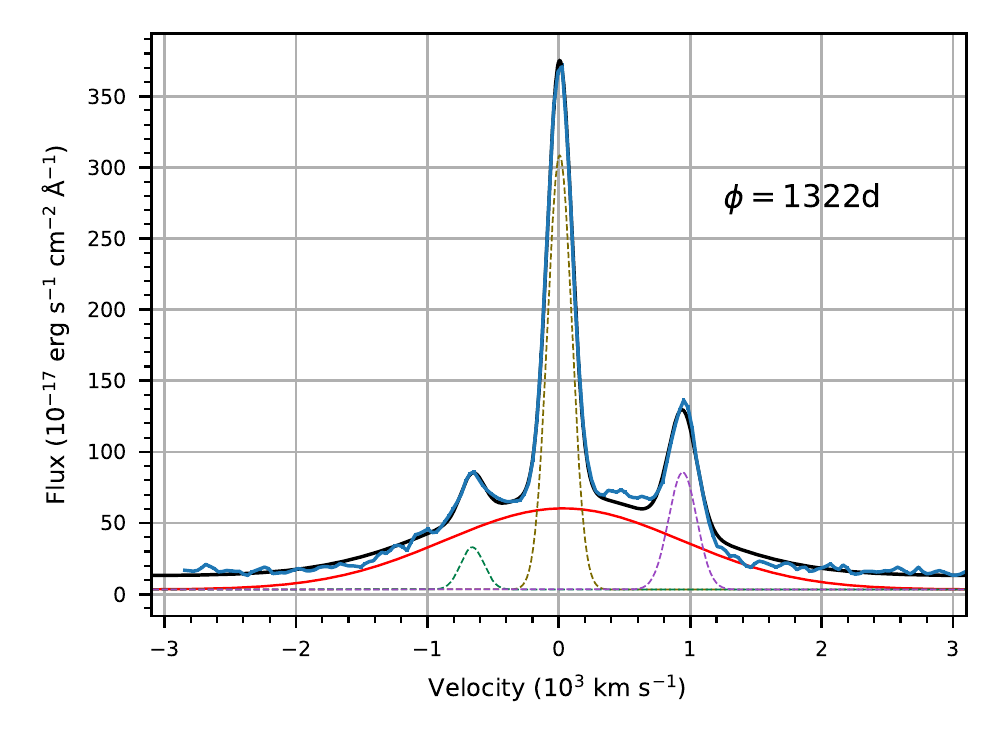}
    \includegraphics[width=\columnwidth]{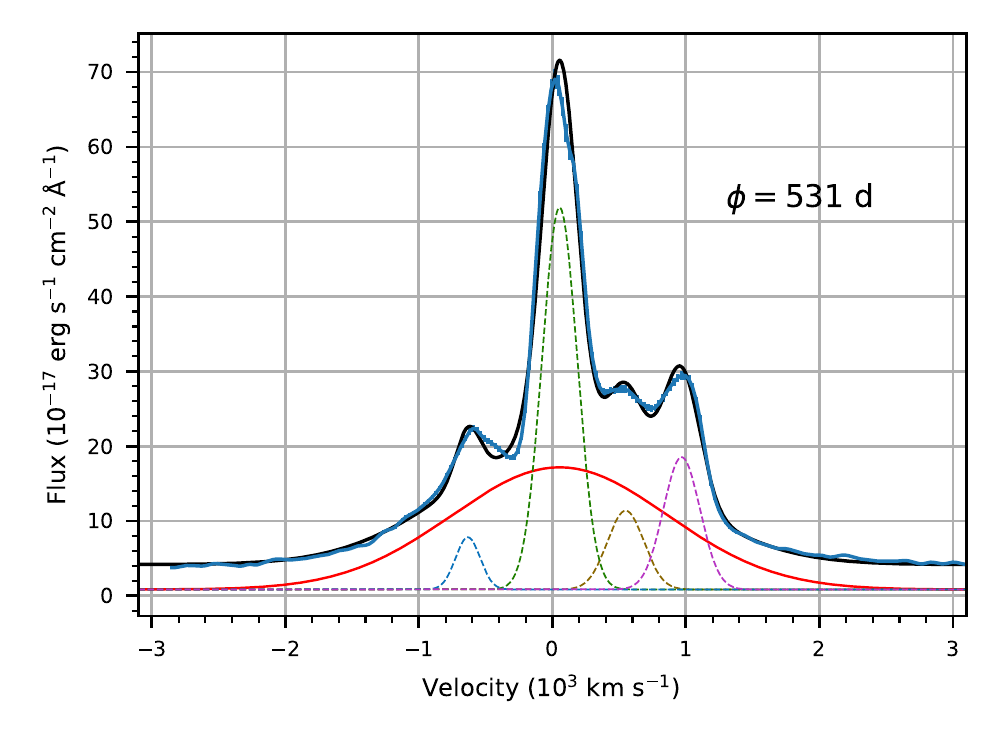}
    \caption{{\it Top panel:} The \Ha\ profile at $\phi = 1322$~d (data in blue) is modelled by the sum of three narrow Gaussian distributions (the two \fnii lines flank the centroid; the narrow \Ha\ line is in gold) plus one additional broad Gaussian that represents the underlying broadened \Ha\ (in red). We determine a FWHM velocity $v = 2120$ \kms\ at $\phi = 1322$~d. {\it Bottom panel:} An additional fifth component is needed to model the data between phases $\phi = 275-623$ d. The fifth component is shown in gold in this fit to public data at $\phi = 531$ d. The centroid of the fifth component moves from blue to red across this phase range. 
    }
    \label{fig:benfit}
\end{figure}

We also analysed publicly available optical spectra downloaded from WISEREP\footnote{\href{https://www.wiserep.org}{https://www.wiserep.org}} \citep{wiserep} to derive the \Ha\ velocity at times that pre-date our earliest HET observation ($\phi = 947$~d) and to look for deviations from the nearly constant \Ha\ velocity that we observe at $\phi > 947$~d. We follow an identical procedure to fit the \Ha\ profile in the  public data as we do for our own spectra. We find that the derived \Ha\ FWHM velocity is essentially constant from $\phi = 127$~d \citep{Milisavljevic15} to the final HET observation at $\phi = 2493$~d.

In addition to measuring the velocities available in the public spectra, we also identified an anomalous additional emission profile within the \Ha\ complex between days $\phi = 275 - 1027$~d \citep{mauerhan18} the central wavelength of which (and hence apparent bulk velocity; see \S \ref{subsec:vel}) appears to redden with time. We note that \cite{Anderson17} included a fifth component in their \Ha\ Gaussian fits to two Keck-II/DEIMOS spectra obtained at $\phi = 530$~d and $\phi = 650$~d that appears to be emitted between the \Ha\ line and the red [N II] line, although they do not offer an interpretation of the additional component. We present our own fit to an example \Ha\ complex containing this additional fifth component in Figure \ref{fig:benfit} (bottom panel).

\begin{figure}
    \centering
    \includegraphics[width=\columnwidth]{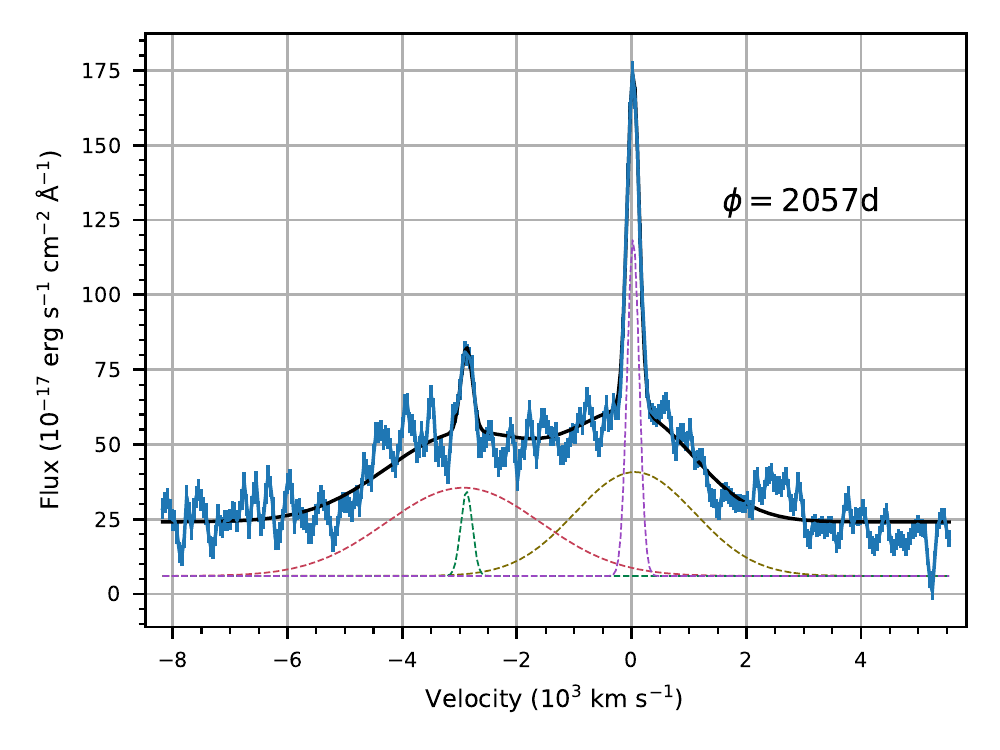}
    \caption{The \foiii\ 4959/5007 profile is modelled with a sum (model in black) of two narrow Gaussian distributions plus two broad Gaussian distributions representing the narrow and broadened components of the emitted \foiii\ flux (data in blue). We determine a FWHM velocity of $v = 2460$ \kms\ and $v = 3190$ \kms\ for the \foiii\ 5007 and 4959 lines, respectively. The velocity derived from FWHM of the \foiii\ remains fairly constant at around $v \approx 3000$ \kms\ throughout the duration of our observations.}
    \label{fig:oiii_fit}
\end{figure}

We follow a similar procedure to fit the \foiii\ $\lambda \lambda 4959, 5007$ complex. We note that both narrow and broad components to this doublet are present in our spectra, but the narrow components tend to fade with time. The widths of the two broad components may have some additional error associated with them due to cross-contamination with the narrow \Hb\ line that is present to the blue of the \foiii\ complex. The \foiii\ complex is modelled as the sum of two narrow Gaussian distributions and two broadened distributions. Again, we are primarily interested in the broadened \foiii\ wings as they are most likely indicative of activity relating to the SN. The fit to our $\phi = 2057$~d spectrum is given in Figure \ref{fig:oiii_fit}. We derive FWHM values of the broadened components at $\phi = 2057$~d of FWHM = $52.8\pm5.09$\AA\ and FWHM = $41.1\pm5.09$\AA\ for \foiii\ $\lambda 4959$ and $\lambda 5007$, respectively. These widths correspond to velocities of ${\sim}3000$ \kms. This velocity remains relatively constant across the duration of our observations. 

By virtue of a similar method, we have also derived line widths (and thereby velocities) for \ion{He}{1} $\lambda 7065$ (FWHM = $96.1\pm4.24$ \AA\ at $\phi = 947$~d), [\ion{O}{1}] $\lambda 6300$ (FWHM = $109.3\pm5.09$ \AA\ at $\phi = 976$~d), [\ion{Ca}{2}] $\lambda 7291, 7324$ (FWHM = $96.2\pm4.24$ and FWHM = $157.2\pm4.24$ \AA, respectively, at $\phi = 947$~d). Each of the above FWHM measurements has a very small error from the fit contribution at $<1\%$, and a dominant error from the spectral resolution at ${\sim}10\%$. We expect an additional uncertainty in the [Ca II] lines due to an  absorption immediately to the blue of the doublet that obfuscates the continuum level (Figure \ref{fig:narrow}). We nonetheless interpret the FWHM of each individual transition as essentially constant across the observed epochs.  We find no evidence for a broad component to the \Hb\ line although such a faint, broad component may be hidden beneath the noise level.

\subsubsection{IR spectra and the He line profile} \label{subsubsec:IR}

\citet{tinyan19} present NIR 1-2.5 \mic\ spectra using TripleSpec on P200 \citep{Herter08}, and the Near-Infrared Echellette Spectrometer (NIRES) and the Multi-Object Spectrometer for Infra-Red Exploration (MOSFIRE) on the Keck telescope. Their spectra span the epochs from $\phi = 282$ to $\phi = 1707$~d (their Figure 4). The data at $\phi = 282$~d do not quite reach as blue as the He I 1.0830 {\mic} line, but show a broad feature of He I 2.058 {\mic}. Data from $\phi = 1319$~d show a very strong broad feature of He I 1.0830 {\mic} and a weaker broad feature of 2.058 {\mic} along with narrow hydrogen lines. There seem to be no detected broad hydrogen features. 

\citet{tinyan19} presented Gaussian decomposition fits of the He I 1.0830 {\mic} line at two epochs ($\phi = 1368$ and 1707~d). Inspection of their Figure 8 shows that the FWHM of the strongest, broadest component (component `a') corresponds to a velocity width of $\gtrsim 4000$ \kms. The He I 2.058 {\mic} line has a comparable width at $\phi = 282$~d, but the line becomes less prominent later. \citet{tinyan19} also identified two lower-amplitude, narrower components that they attribute to the He I 1.083 {\mic} line, one centered at a blueshift of -4000 \kms\ (component `b') and one centered near zero velocity (component `c').  Finally, there is a narrow unresolved but relatively strong 1.083 {\mic} line centered at rest and a narrow unresolved H I 1.094 {\mic} line presumably also in the same rest frame as the narrow He I component.

\begin{figure}
    \centering
    \includegraphics[width=\columnwidth]{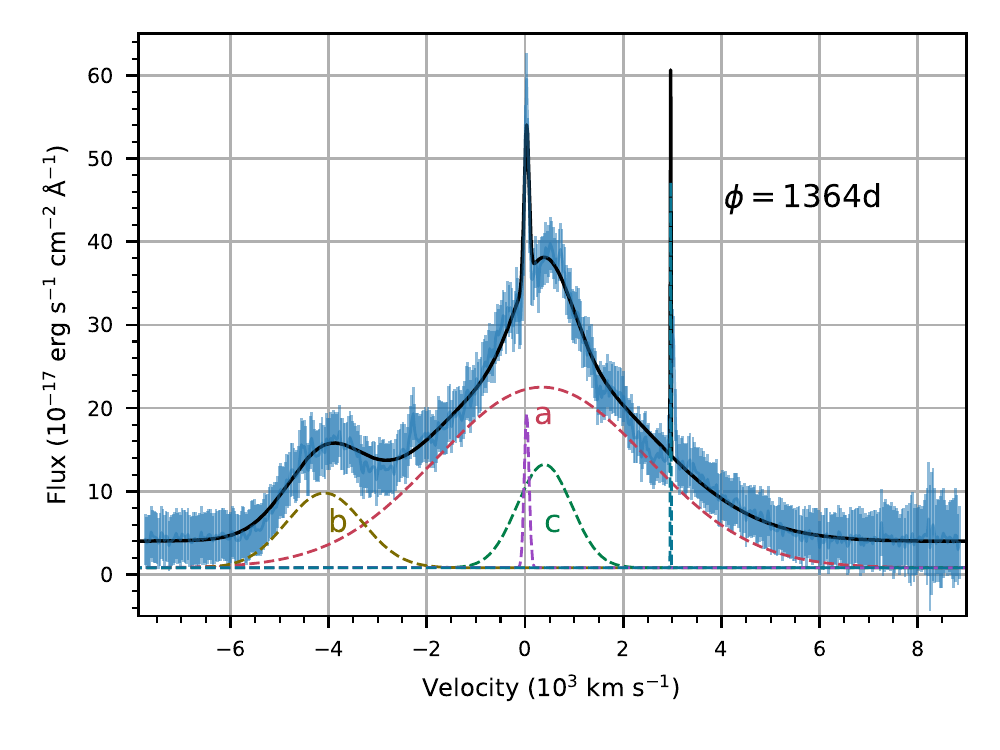}
    \caption{The \ion{He}{1} 10830 \AA\ profile from the $\phi = 1364$~d data of \citet{tinyan19} is modelled with a sum (model in black) of a broad Gaussian (component `a'), two narrower, weaker components (`b' and `c'), an unresolved narrow line centered at rest, and a narrow unresolved H I 1.094 \mic\ line. }
    \label{fig:hei_fit}
\end{figure}

We have performed our own multiple Gaussian fit to the 1.083 {\mic} line following the procedures outlined in \S \ref{subsubsec:profiles} and as illustrated in Figure \ref{fig:hei_fit}. We find FWHM = $182.7\pm4.73$ \AA\ at $\phi = 1364$~d for the broad, central `a' component. At the same epoch, but for the narrower sub-components we find FWHM = $66.2\pm4.73$ \AA\ for component `b' that is centered at -4000 \kms\ and FWHM = $47.1\pm4.73$ \AA\ for rest component `c'.  
In velocity space, these FWHM values correspond to $5050\pm130$, $1860\pm130$ and $1300\pm130$ \kms\ for components `a', `b' and `c', respectively.
In the absence of access to flux-calibrated spectra, we are unable to estimate NIR helium line fluxes or luminosities.

\subsubsection{Constraints from Line Profiles} \label{subsub:profiles}

Asymmetries and aspect angle effects could play a role in \sn\ with implications for the line profiles of the broadened lines we observe. The narrow lines are not resolved, so are probably not affected. The broader emission lines could give evidence for the distribution of composition and density of gas and dust, and for the aspect angle of the observer.

 If the \Ha\ is associated with an expanding ring of emission with a hole in the center that suppresses flux at low velocity, the line is expected to show two ``horns" symmetrically displaced around the line center \citep{jerk17}. Such a profile is not obvious, but may not be ruled out. It is difficult precisely to determine the profiles of the \Ha\ emission lines that we and others have observed because of the presence of the strong narrow component, the two [\ion{N}{2}] lines that straddle \Ha\, and convolution with the instrumental resolution that will smooth out any complex substructure to the broadened profile.  To the extent that the \Ha\ profiles do not show the expected double peak, the observations are inconsistent with a model for the \Ha\ emission based on a simple thin shell expanding at constant velocity.
 
As shown in Figure \ref{fig:benfit}, the principle Gaussian that matches the broad wings of \Ha\ is centered on the narrow feature at zero velocity. In principle, this puts a limit on any dust extinction or non-axisymmetric distribution of the emitting hydrogen due to basic geometry, as is seen in some SN~IIn \citep{Smith11iqb}. 

The \Ha\ lines do show the odd ``travelling fifth component" at some phases (Figure \ref{fig:benfit}; \S \ref{subsec:vel}) that appeared to shift redward between $\phi = 275 - 531$~d. The timescale of the drift of this feature is about right for an orbital period of ${\sim}300$~d \citep{sunmc20}, but the velocity displacement (-420 to + 540 \kms; Table \ref{tab:fwhm_ha5}) and the width of the Gaussian fit (FWHM $\sim 300$ \kms) are too large to correspond to expected orbital motion of any neutron star or companion, ${\sim}10$ \kms \citep{sunmc20}. It is conceivable that a pulsar in an eccentric orbit blowing a fast wind could contribute to such a feature.

The peak of the main Gaussian and that of component `c' in the \ion{He}{1} 1.083 \mic\ line in Figure \ref{fig:hei_fit} are each displaced to the red by 338 and 410 \kms, respectively. This displacement is the opposite of that expected for dust obscuration and is in marked contrast to the lack of any such displacement of \Ha. The red displacment of the \ion{He}{1} might be due to some non-axially symmetric dynamic effect from the formation of the CSM \citep{Smith11iqb}. An alternative is that we are seeing emission from the material of the helium-rich wind that is ``behind" the reverse shock and hence heading away from us on the near side of the structure. A corresponding blue-shifted component on the far side might be obscured by dust or by the dense SN ejecta itself. Other alternatives for this red displacement are the result of the interaction of the fast helium wind of the progenitor star with that of the main sequence companion or of some asymmetry in the explosion that specifically affects the helium distribution and excitation. 

Component `b' of the \ion{He}{1} 1.083 \mic\ line is displaced to the blue by 4076 \kms. The FWHM of the corresponding Gaussian fit to this sub-feature is ${\sim}1859$~\kms. The lack of any such component to the red could be due to dust obscuration or to an intrinsic departure from axisymmetry. It is possible that component `b' is just a separate small emission feature unrelated to \ion{He}{1}. No feature with a displacement of ${\sim}4000$~\kms\ is associated with the \Ha\ line, but such a feature could be confused with the emission line of [\ion{O}{1}]. A careful check suggests that such a hypothesized feature would be too red by about $3\sigma$ to overlap with [\ion{O}{1}].

The emission features of [\ion{O}{3}] $\lambda\lambda$ 4959, 5007 are reasonably well fit by single Gaussians as shown in Figure \ref{fig:oiii_fit}. There is no evidence of double peaks. Figure \ref{fig:oiii_fit} shows that the 5007 line is closely centered on zero velocity. The profiles are consistent with emission from a filled volume as would be expected from the inner ejecta. Although it is likely that the oxygen emission comes from the ejecta as does the helium emission, the oxygen may show no red/blue asymmetry because it occupies a smaller volume that is less susceptible to differential extinction. The relatively high excitation features might be related to a central pulsar.

\subsubsection{Narrow Lines} \label{subsubsec:narrow}

\begin{figure*}
    \centering
    \includegraphics[width=\textwidth]{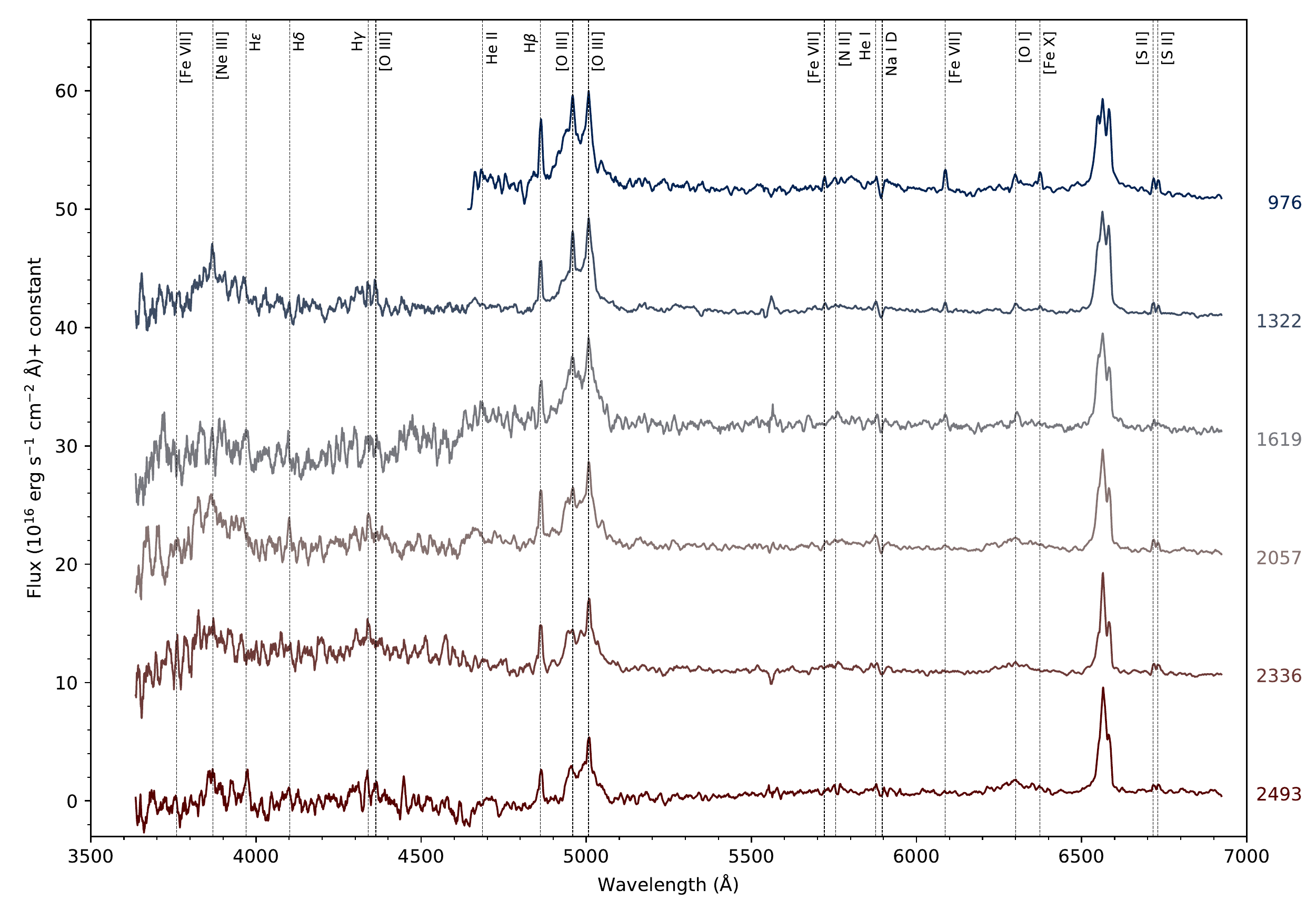}
    \includegraphics[width=\textwidth]{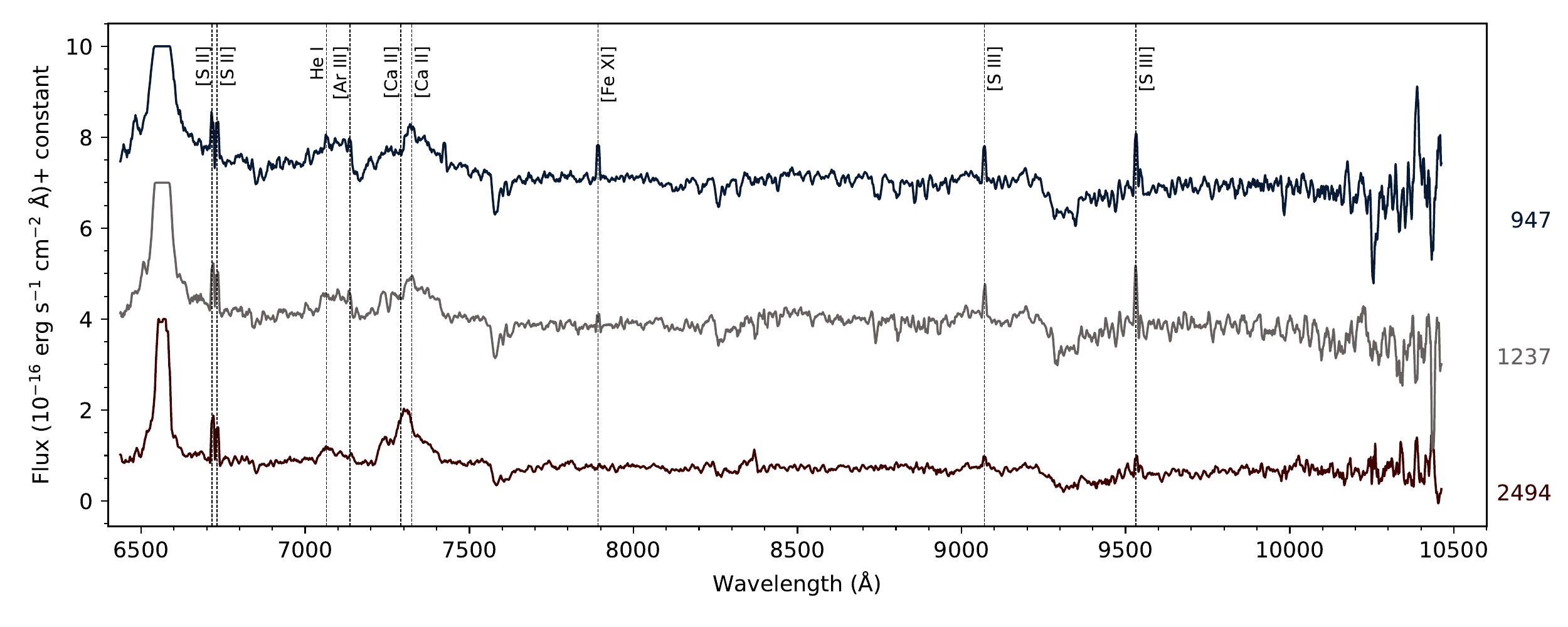}
    \caption{Line identifications for six LRS2-B spectra ({\it top}) and three LRS2-R spectra ({\it bottom}). A 15 \AA\ smoothing kernel has been applied to tease out faint, broad components. The dominant \Ha\ and [\ion{O}{3}] profiles have been clipped for clarity. In addition to broadened emission from Ha, [\ion{O}{3}]\ $\lambda\lambda 4959, 5007$ and \foi\ $\lambda 6300$,  there are discernible broadened components to the [\ion{Ne}{3}] line (the narrow component of which fades entirely between $\phi = 1322$ and 1619~d), the H$\gamma$/[\ion{O}{3}]\ $\lambda 4363$ doublet and \ion{He}{2}\ $\lambda 4686$. }
    \label{fig:narrow}
\end{figure*}

Narrow line identifications are presented in Figure \ref{fig:narrow}. The FWHM values are quantified using Gaussian fits to the narrow emission lines. The uncertainty from the fit on the FWHM values is around $0.1$ \AA\ (or ${\sim}2.5\%$), while the uncertainty contribution from the spectral resolution is again expected to be much larger ($4-5$ \AA). Our spectrum taken at $\phi = 1322$~d, which has the best signal-to-noise ratio, is used to measure the majority of lines blueward of \fsii (exclusive). Our spectrum taken at $\phi = 947$~d is used to identify and measure lines redward of \fsii (inclusive). 

We also consider FWHM values as measured from our background spectra defined as flux from the area indicated by the grey squares in Figure \ref{fig:IFUimage}. Reductions of the data from the background and the supernova are performed on the same total IFU image and therefore must suffer from the same weather limitations. Direct comparisons between the supernova and background line widths are therefore useful to determine whether or not the narrow supernova lines are resolved. The background lines are presumed to be unresolved, and thus their measured FWHM values give an indication of the instrumental resolution. In all cases, the narrow emission line widths from the background spectrum are comparable to or broader than the corresponding lines in the supernova spectrum. From hereon we do not consider the width of the narrow lines from the supernova to be meaningful within an astrophysical context (although their other properties, such as integrated fluxes, may still be meaningful).

\citet{Kim14} determine the redshift of the host (NGC 7331) to be $z = 0.002722$. We determine a redshift from the narrow lines of our spectra of $z = 0.003175$; an additional redshift with respect to the host. The implied SN velocity relative to the host is 136 \kms, fairly typical of galaxy spin velocities \citep{Sofue01}. We interpret this as evidence that \sn\ is in an arm of the host galaxy with a velocity whose radial component points away from the observer. 

\subsubsection{Information From Line Ratios} \label{subsubsec:ratio}

\citep{osterbrockF} give the electron temperature as a function of the [\ion{O}{3}] (5007 + 4959) / 4363 narrow line ratio. Their estimation of the electron temperature from the [O III] emission lines depends upon a low density approximation, where the electron density must be $n_{\rm e} < 10^5$ cm$^{-3}$, above which the lower energy 4959, 5007 lines begin to get collisionally de-excited. 

We detect the relatively weak [\ion{O}{3}] 4363 line at $\phi = 1322$~d along with the stronger [\ion{O}{3}] 4959, 5007 lines. Using narrow Gaussian fits, we compute a flux ratio [\ion{O}{3}] (5007+4959)/4363 = 5.06. From \cite{osterbrockF}, this may imply a lower bound to the temperature of $T > 20,000$ K; however, if these lines originate from the inner ejecta, it may be that the 4959, 5007 transitions are collisionally de-excited, at which point this approximation would break down. It is thus difficult to distinguish between the possibility of radiative versus collisional deexcitation and hence to determine a temperature from [\ion{O}{3}].

The narrow lines of \fsii\ $\lambda \lambda$6716, 6731 are clearly detected in all of our spectra (Figures \ref{fig:narrow} and \ref{fig:snhII2}). The ratio of these lines gives a measure of the density \citep{osterbrockF}. We find that the line strength varies gradually with time, but that the line ratio is essentially constant $\approx 1.2$. This gives a density ${\sim}100$~cm$^{-3}$, much less than that determined from, e.g., X-ray emission \citep{Margutti17}. The fact that the \fsii\ maintains the same line ratio and hence density means that the material radiating the lines must be essentially static on the timescales involved. The fact that the \fsii\ flux varies in time suggests that it is somehow exposed to photoionizing flux, if non-locally. 

Our spectra show no evidence of \fsii\ $\lambda \lambda$4068, 4072 that might provide a constraint on density in comparison with \fsii\ $\lambda \lambda$6716, 6731.

\subsubsection{Spectra of the Environment} \label{subsubsec:environment}

\begin{figure*}
    \centering
    \includegraphics[width=.8\textwidth]{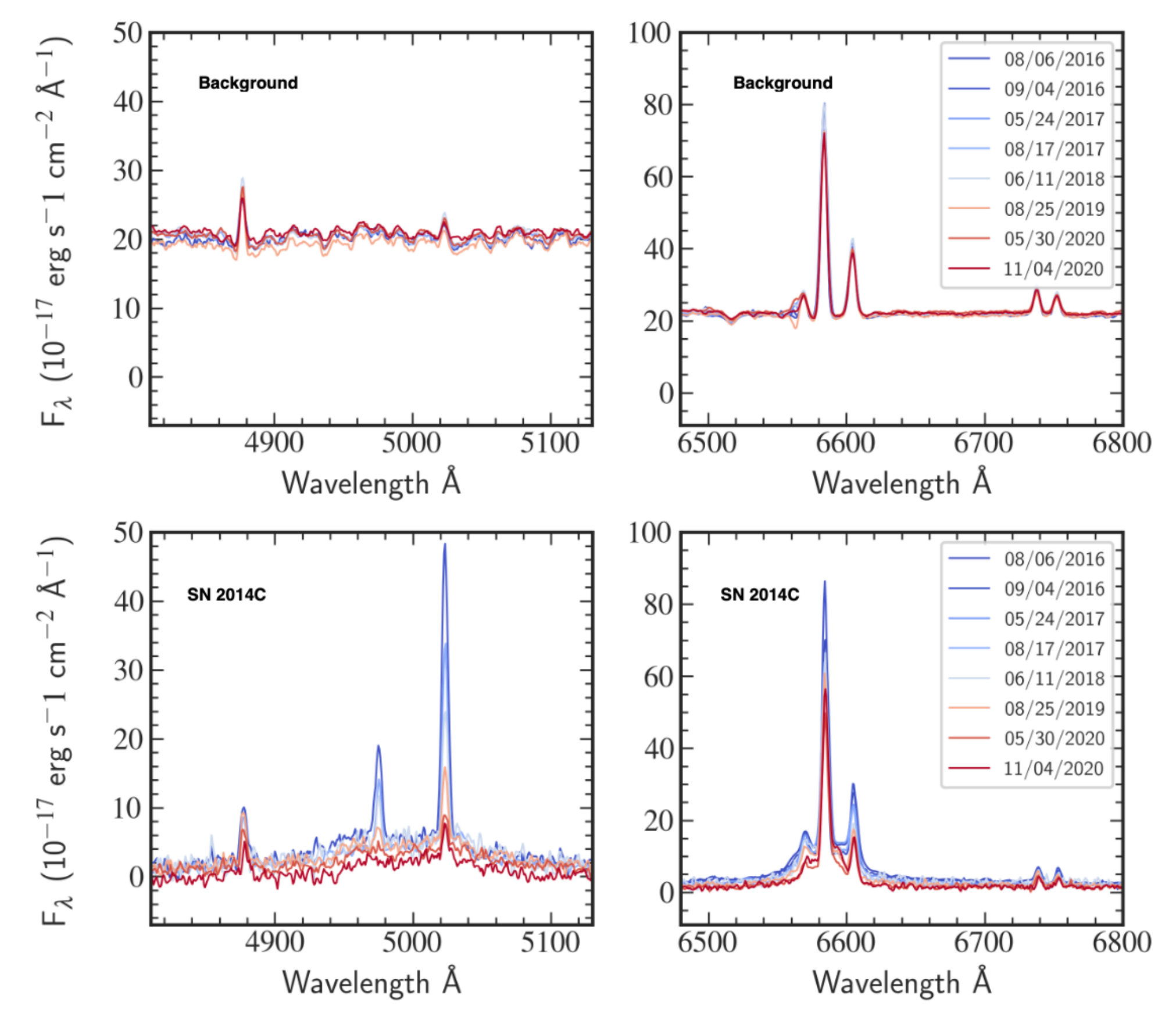}
    \caption{The time evolution is presented for LRS2 IFU spectra of the spatially resolved H II region (top panels) and the location of SN~2014C (bottom panels) for the wavelength region encompassing \Hb\ and \foiii (left) and that covering \Ha, \fnii and \fsii(right). Note that the continuum and narrow line emission from the H II region are basically constant in amplitude and emission line width while those from the vicinity of the supernova seem to decline in strength.  The spectra were flux-calibrated to the galaxy background emission in LRS2 (not shown here) and the consistency of the H II region spectra with time translates to a quantification of flux calibration.  The implication is that the variations seen in the SN~2014C spectrum in the bottom panels are real and not an artifact of calibration.}
    \label{fig:snhII2}
\end{figure*}

Our IFU spectra give us the opportunity to compare the spectrum from the location of SN~2014C with that of nearby locations in the host galaxy. Figure \ref{fig:snhII2} gives a comparison of the supernova environment with that of a knot in the nearby spatially-resolved H II region revealed in Figures \ref{fig:image} and \ref{fig:IFUimage} as a function of epoch for our nine spectra. Spectra are shown for the wavelength region around \Hb and \foiii and around \Ha, \fnii and \fsii.

From Figure \ref{fig:IFUimage}, the separation of the knot in the H II region and \sn\ is about 2.15 arcseconds. The observational seeing ranges from 1.6-3.0 arcseconds, which can be larger than the SN - H II region separation. A distance of 14.7 Mpc would imply a separation of 150 pc, probably too far for the SN to irradiate the H II region and cause it to emit in \Ha.  The galactic background spectra are obtained from the median spectrum from within the gray square in Figure \ref{fig:IFUimage}.  The black boundary is the fitting region for the point source models of the H II knot and SN sources.  

The narrow lines from the H II region shown in Figure \ref{fig:snhII2} are basically constant in amplitude and width with any variation attributable to variations in observing conditions such as air mass and seeing. In contrast, the narrow lines from the vicinity of the supernova seem to decrease in strength by about a factor of two from the early to the later spectra for the \Ha, \Hb, \fnii, and \fsii, and closer to a factor of 10 for the narrow \foiii lines. While it is possible that the latter variation might also be attributed to observing conditions, it seems to be systematic in time suggesting that whatever is hosting those narrow lines is itself subject to irradiation by the supernova. 

The density associated with the \fsii\ lines, ${\sim}100$~cm$^{-3}$, is characteristic of an H II region. The narrow \fsii\ lines could thus be associated with an unresolved nearby (less than 1 arcsecond ${\sim}75$ pc) ambient H II region that has nothing directly to do with \sn\ but could be irradiated by it. An alternative is that the narrow \fsii\ lines could arise from the low density outer reaches of the CSM expelled by the supernova progenitor system. Note that the narrow lines reported in \citet{Milisavljevic15} may be a convolution of emission from the constant and the putative variable H II region.

\section{X-rays} \label{sec:xray}

\subsection{Data Reduction and Spectral fitting} \label{sec:reduction}
SN~2014C was first detected in the X-ray band by Swift/XRT on 2014 January 6th (phase $\phi = 6$ d), followed by a series of observations with a 13 day cadence. The source then entered solar occlusion and the next X-ray observations were obtained in November 2014. We have reduced all the data taken by \chandra\ and \nustar\ between 2014 November and 2020 April, as well as the  \swift\ observations. The data were all downloaded from the respective satellite archives. All but the first \chandra\ observation were coordinated with \nustar, providing coverage over a broad energy range. Table~\ref{tab:obslog} gives the detailed log of the available data and the derived key parameters. All data were reduced according to the standard reduction procedures of each satellite. We discuss the data reduction and analysis in detail below. 

\begin{deluxetable*}{lcccccccccc}[ht!]
\rotate
\tablecaption{Summary of X-ray data on \sn, listed in chronological order, including the satellite that performed the observation, the observation date, the days after explosion, the exposure time, the column density, derived temperature, iron abundance and unabsorbed luminosity. A subset of these data may also be found in \cite{Margutti17, Brethauer20}. \label{tab:obslog}}
\tablecolumns{11}
\tablewidth{0pt}
\tablehead{
\colhead{Satellite} & \colhead{Obs. date} & \colhead{ Obs. ID} & \colhead{ PI} &
\colhead{Age}  & \colhead{Exposure}& \colhead{Count rate} & \colhead{$N_\mathrm{H}$} &  \colhead{kT}&\colhead{\Afe}& \colhead{$L_X^\mathrm{0.3-100\,keV}$} \\
& & & &(days)&\colhead{(ks)} & \colhead{($10^{-3} \mathrm{counts}\,\mathrm{s}^{-1}$)} & \colhead{($10^{22}\,\mathrm{cm}^{-2}$)} & \colhead{(keV)} &&\colhead{($10^{40}\,\mathrm{erg}\,\mathrm{s}^{-1}$)}
}
\startdata
\swift & 2014-01-(06 to 19) & 000330780(01)-(20) & Milne & 7-20 & 17.5 & $8.45\times10^{-4}$ & $\--^a$ & $\--^a$ & $\--^a$ & $<0.48^b$ \\
\hline
\chandra & 2014-11-03 & 10.25574/16005 & Soderberg & 308 & 9.9 & $1.22\times10^{-2}$ & $5.20^{+2.93}_{-1.98}$ & $>25.03$ & $>5.11$ & $3.28^{+0.51}_{-0.51}$$^{c}$\\
\hline
\nustar & 2015-01-29 & 80001085002 &  Margutti & 395 & 32.5 & $2.57\times10^{-2}$ & \multirow{2}{*}{$3.75^{+0.91}_{-0.76}$} & \multirow{2}*{$12.5^{+3.0}_{-2.2}$}  & \multirow{2}*{$3.25^{+1.71}_{-1.06}$} & \multirow{2}*{$4.95^{+0.43}_{-0.43}$}\\
\chandra & 2015-01-30 & 10.25574/17569 & Margutti & 396 & 9.9 & $2.26\times10^{-2}$ & ~ & ~ & ~ \\
\hline
\nustar & 2015-04-19 &  40102014001 &  Margutti & 475 & 22.4 & $2.47\times10^{-2}$ & \multirow{2}{*}{$3.32^{+1.00}_{-0.81}$} & \multirow{2}*{$14.8^{+4.3}_{-3.3}$} & \multirow{2}*{$4.74^{+3.29}_{-1.91}$} & \multirow{2}*{$5.46^{+0.50}_{-0.50}$}\\
\chandra & 2015-04-20 &  10.25574/17570 & Margutti & 476 & 9.9 & $2.64\times10^{-2}$ & ~ & ~ & ~ \\
\hline
\chandra & 2015-08-28 & 10.25574/17571 & Margutti & 606 & 9.9 & $2.56\times10^{-2}$ & \multirow{2}{*}{$1.93^{+0.59}_{-0.55}$} & \multirow{2}*{$13.4^{+5.5}_{-1.9}$} & \multirow{2}*{$3.81^{+2.73}_{-1.11}$} & \multirow{2}*{$5.44^{+0.45}_{-0.45}$}\\
\nustar & 2015-08-29 & 40102014003 & Margutti & 607 & 30.2 & $3.12\times10^{-2}$ & ~ & ~ & ~ \\
\hline
\nustar & 2016-05-03 & 40202013002 & Margutti & 855 & 43.0 & $2.66\times10^{-2}$ & \multirow{2}{*}{$1.18^{+0.21}_{-0.19}$} & \multirow{2}*{$11.5^{+1.6}_{-1.6}$} & \multirow{2}*{$2.35^{+0.75}_{-0.62}$} & \multirow{2}*{$5.48^{+0.30}_{-0.30}$}\\
\chandra & 2016-05-05 & 10.25574/18340 & Margutti & 857 & 27.7 & $4.56\times10^{-2}$ & ~ & ~ & ~ \\
\hline
\chandra & 2016-10-24 & 10.25574/18341 & Margutti  & 1029 & 29.6 & $4.98\times10^{-2}$ & \multirow{2}{*}{$0.93^{+0.14}_{-0.13}$} & \multirow{2}*{$11.8^{+1.5}_{-1.5}$} & \multirow{2}*{$3.69^{+1.00}_{-0.79}$} & \multirow{2}*{$5.72^{+0.31}_{-0.31}$}\\
\nustar & 2016-11-01 & 40202013004 & Margutti & 1037 & 40.9 & $2.83\times10^{-2}$ & ~ & ~ & ~ \\
\hline
\chandra & 2017-06-09 & 10.25574/18342 & Margutti & 1257 & 28.1 & $5.19\times10^{-2}$ & \multirow{2}{*}{$0.57^{+0.14}_{-0.13}$} & \multirow{2}*{$12.2^{+2.2}_{-1.8}$} & \multirow{2}*{$4.26^{+1.47}_{-1.12}$} & \multirow{2}*{$4.85^{+0.30}_{-0.30}$}\\
\nustar & 2017-06-16 & 40302002002 & Margutti & 1264 & 42.3 & $2.16\times10^{-2}$ & ~ & ~ & ~ \\
\hline
\chandra & 2018-04-16 & 10.25574/21077 & Margutti & 1568 & 19.8 & $5.39\times10^{-2}$ & \multirow{3}{*}{$0.52^{+0.14}_{-0.13}$} & \multirow{3}*{$10.2^{+1.6}_{-1.1}$} & \multirow{3}*{$2.36^{+0.70}_{-0.58}$}  & \multirow{3}*{$4.68^{+0.26}_{-0.26}$}\\
\chandra & 2018-04-22 & 10.25574/18343 & Margutti & 1574 & 9.9 & $4.99\times10^{-2}$ & ~ & ~ & ~ \\
\nustar & 2018-05-04 & 40302002004 & Margutti & 1586 & 40.2 & $2.25\times10^{-2}$ & ~ & ~ & ~ \\
\hline
\chandra & 2019-05-24 & 10.25574/21639 & Margutti & 1971 & 29.5 & $4.29\times10^{-2}$ & \multirow{2}{*}{$0.38^{+0.12}_{-0.11}$} & \multirow{2}*{$8.2^{+1.3}_{-1.0}$} & \multirow{2}*{$1.94^{+0.63}_{-0.49}$} & \multirow{2}*{$3.47^{+0.22}_{-0.22}$}\\
\nustar & 2019-06-01 & 40502001002 & Margutti & 1979 & 44.5 & $1.90\times10^{-2}$ & ~ & ~ & ~ \\
\hline
\chandra & 2020-04-16 & 10.25574/21640 & Margutti & 2299 & 17.8 & $3.63\times10^{-2}$ & \multirow{3}*{$0.24^{+0.14}_{-0.13}$} & \multirow{3}*{$8.3^{+1.3}_{-1.0}$} & \multirow{3}*{$1.73^{+0.54}_{-0.43}$}  & \multirow{3}*{$2.21^{+0.14}_{-0.14}$}\\
\chandra & 2020-04-18 & 10.25574/23216 & Margutti & 2301 & 10.9 & $3.85\times10^{-2}$ & ~ & ~ & ~ \\
\nustar & 2020-04-30 & 40502001004 & Margutti & 2313 & 54.2 & $1.56\times10^{-2}$ & ~ & ~ & ~ \\
\enddata
\tablenotetext{a}{The parameter cannot be derived due to low counts but is estimated by \pimms.}
\tablenotetext{b}{The luminosity is estimated in \S \ref{subsec:swift} using 0.3-10.0 keV \swift\ observations.}
\tablenotetext{c}{The luminosity is corrected based on the later observations as explained in \S \ref{subsec:CXO-NuSTAR}.}

\end{deluxetable*}

\subsection{\swift\ Extraction} \label{subsec:swift}
The {\it Neil Gehrels Swift Observatory} consists of the Burst Alert Telescope (BAT),	X-ray Telescope (XRT) and Ultraviolet/Optical Telescope (UVOT) \citep{2005Burrows}. We only included \swift/XRT data covering the phase $\phi \sim7\mbox{--}20$ days after first optical light when conducting the analysis. The \swift\ data were reduced following the standard procedures using \swift\ Data Analysis Software (XRTDAS v0.13.5) and updated XRT calibration files \caldb\ (v20190910). We produced the calibrated and filtered event files with the \xrtpipeline\ script. All of these event files were combined using the \xselect\ package. We extracted the source spectrum from a circular region of 10-arcsec radius centered on the source (position information obtained from SIMBAD Database\footnote{http://simbad.u-strasbg.fr/simbad/}), and the background spectra from an identical circular region away from the source. At the early epochs, the observed X-ray counts were not sufficient to allow spectral fitting. Instead, we estimated the upper limit to the flux using the Bayesian method proposed by \citet{1991Kraft}. There are 9 photons within 10 arcsec, of which 4 are expected to be from the background. Using \citet{1991Kraft}, we derive a 99-percent confidence level of the upper limit of 54.4 counts, and thereby a count rate of $8.45\times10^{-4}~\mathrm{c~s^{-1}}$ (0.3-10 keV), given the 17.5~ks exposure time. The unabsorbed flux is obtained by inputting this count rate into \chandra\ \pimms (v4.10), to deduce an upper limit of $1.86\times10^{-13}$ erg cm$^{-2}$ s$^{-1}$ assuming an absorbed thermal model. The hydrogen column density and the temperature were fixed to the values obtained from the first \chandra\ observation, \nh$=5\times10^{22}~\mathrm{cm}^{-2}$ and kT~$\sim25$~keV, respectively  (Table.\ref{tab:obslog}). The corresponding upper limit to the luminosity is $4.79\times10^{39}$ erg s$^{-1}$.

\subsection{Chandra Extraction}\label{subsec:chandra}
The spatial and spectral resolution of the \chandra\ X-ray Observatory \citep[CXO;][]{2002Weisskopf} allows the position and emission lines of \sn\ to be resolved. Chandra observations were performed with the Advanced CCD Imaging Spectrometer S-array (ACIS-S) instrument on \chandra, starting from 2014 November ($\phi = 35$ d). The \chandra\ analysis was done using \ciao\ (v4.11) software and corresponding calibration files. The data are reprocessed, and the source spectra extracted from a 3-arcsec region centered on \sn. Background spectra are extracted from an 8-arcsec source-free region, and subtracted from the source. Response files (ARF and RMF) are created using \specextract . \chandra\ spectra of \sn\ are available at twelve different epochs. At two epochs, in April 2018 and April 2020, the observations were taken less than a week apart, and are therefore combined together, using the \combine\ script. 

\subsection{\nustar\ Extraction}\label{subsec:nustar}
The {\it Nuclear Spectroscopic Telescope Array} \citep[\nustar;][]{2013Harrison} is the first space-based satellite focusing on the hard X-ray band from 3 to 79 keV.

\sn\ was observed by the FPMA/B instruments nine times between 2015 and 2020. The \nustar\ data were processed with the \nustar\ Data Analysis Software (NUSTARDAS v.1.8.0) and the calibration files in \nustar\ CALDB (v20190812). We use the \nupipeline\ package to create calibrated event files. Both the source and background spectra are extracted from a 1-arcmin circular region. Due to the poor angular resolution compared to \chandra, the \nustar\ spectra are contaminated by emission from nearby objects. The spectra, response matrix files, and position-dependent ancillary response files are generated by using the \nuproducts\ program.

\subsection{Spectral fitting} \label{subsec:CXO-NuSTAR}

Chandra\ covers the energy range of $0.3\mbox{--}10$\,keV with a point-spread function (PSF) of $0.5''$ FWHM, which is able to spatially resolve \sn\ from other X-ray sources in its host galaxy NGC7331, while \nustar\ is effective between 3 and 79\,keV with a wider PSF of $18''$ FWHM. The latter cannot easily resolve the supernova, and contamination from other sources in the $1'$ extraction region was a concern. To estimate the degree of contamination, we follow \citet{Margutti17}. We extract \chandra\ spectra of the contaminated region from an annular region with inner radius $3''$ and outer radius $1'$ centered on \sn. The spectra are fitted by an absorbed power-law model, and the derived spectral parameter values are interpolated into the \nustar\ spectral fitting by adding a background component. We found that the additional background component did not make a significant contribution to the spectra. This can be understood since most of the emission from this component is at an energy lower than 3\,keV, which is below the \nustar\ energy range. \chandra\ spectra are grouped to have at least 15 counts in each bin, while the \nustar\ data are grouped to 20 counts to have sufficiently high signal-to-noise ratio. Given sufficient counts in each observation to allow for spectral fitting, the derived parameters are calculated using the $\chi^2$ statistic, with parameter uncertainties estimated at a $90\%$ confidence level.

We analyze \chandra\ and \nustar\ spectra at each epoch simultaneously, with the exception of the first \chandra\ observation, which was not accompanied by a \nustar\ observation. The spectral fitting is carried out using the \xspec\ (v12.10.1f) package \citep{1996Arnaud}, with a thermal emission model. Here we implement fits with the \vapec\ model, which describes the emission from the collisionally-ionized diffuse gas. The \vapec\ model is characterized by temperature $kT$, and the abundance of individual elements. The absorption component is described by the \tbabs\ model \citep{2000Wilms}, characterized by the column density, \nh. The \vapec\ model assumes ionization equilibrium. Ionization equilibrium generally does not hold for young supernova remnants evolving in a low density medium. In that environment, the shock heating causes an abrupt rise in the post-shock temperature, whereas the ionization temperature of the plasma lags far behind and takes time to reach equilibrium with the shock temperature. Ionization equilibrium is roughly reached when the product $n_e t = 10^{12}$ s cm$^{-3}$ \citep{sh10}, where $n_e$ is the gas electron number density and $t$ the time elapsed since the shock impact. Since all the combined  \chandra\ and \nustar\ observations occur after the shock has collided with the high density torus (\S \ref{sec:syn}), the density of which is of order 10$^5$ cm$^{-3}$ \citep[][and \S \ref{sec:syn}]{Margutti17}, ionization equilibrium will be reached in a few months or less, and thus the assumption of ionization equilibrium in the shocked plasma is valid.  

An obvious Fe K$\alpha$ line appears in the \nustar\ spectra. This suggests that the emission is thermal, and Fe may be overabundant. We define the parameter \Afe\ to be the ratio of the mass fraction of iron in the supernova to that in the Sun, with the solar value adopted from \citet{anders91}. We allow \Afe\ to deviate from the unity. A super-solar iron abundance is found (Table \ref{tab:obslog}) that improves the fits by at least $\Delta\chi^2\sim10$ . 

\citet{Margutti17} used an absorbed Bremsstrahlung model to fit the continuum spectra, and then fitted the Fe line separately with a Gaussian. A single absorbed \vapec\ model with variable Fe abundance accomplishes this much more efficiently, with the added benefit that the fitting parameters for both the line and continuum are obtained from a single fit. These differences in spectral fitting lead to small but discernable differences between the flux values derived by us and those of \citet{Margutti17} at the first 4 epochs. This is most noticeable at the epoch of 476 days, where in our case the flux continues to increase compared to the previous epoch of 396 days, whereas in their case the flux decreases from 396 to 476 days. It is difficult to compare the exact values, since \citet{Margutti17} do not provide a table of values of the luminosity. Reading off the value from their light curve plot, it appears there is a difference of only 25-30\% between the flux values at 476 days, which is not a cause for concern.

The derived parameter values, \nh, kT, and \Afe, are listed in Table~\ref{tab:obslog}.  The unabsorbed flux at each epoch is computed using the \cflux\ model in XSPEC. The corresponding unabsorbed luminosity at each epoch is also given in Table~\ref{tab:obslog}. It should be emphasized that for \cxonustar\ fits, we use the \chandra\ data to calculate the flux of $0.3\mbox{--}5.0$\,keV and \nustar\ data to estimate the flux of $5\mbox{--}100$\,keV, because the effective area of \chandra\ begins to decrease as the energy exceeds 5\,keV, while that of \nustar\ starts to decline below 5\,keV. 

The first \chandra\ observation was not accompanied by a contemporaneous \nustar\ observation. In order to calculate the corresponding luminosity over the 0.3-100 keV range, we calculate the contribution of the \chandra\ luminosity to each observation and compute the mean value of the ratio of the \chandra\ luminosity to the total luminosity, which turns out to be ${\sim}50\%$. The first \chandra\ observation is assumed to contribute that same percentage to the total luminosity, thus allowing us to estimate the broad-band luminosity at the epoch of the first \chandra\ observation.

Overall, our analysis shows that the broadband X-ray emission starts to increase from the very first \chandra\ observation, as found by \citet{Margutti17}. The emission continues to increase until just over $\phi = 1000$~d, but then begins to decrease in time. This is different from \citet{Margutti17}, who assumed that the emission decreased after 500 days. The inference is that either \sn\ continues to encounter a high density medium, or that the high level of X-ray emission is being maintained by a different X-ray emission component. The X-ray temperature is highest at 308 days ($>$ 25 keV) and decreases thereafter. Given the error bars, the temperature could also be nearly constant at ${\sim}10$ keV from about 395 days onwards. The iron abundance exceeds solar at all epochs, up to almost 5 times solar at $\phi = 475$~d, but varies epoch-to-epoch. The column density is extremely high at the early epochs, $> 5 \times 10^{22}$ cm$^{-2}$ at an age of $\phi = 308$~d, but decreases steadily thereafter.

\subsection{Constraints from X-rays} \label{subsec:xconstraint}

Observations summarized here and in Table \ref{tab:obslog} showed that the X-ray flux rose quickly for the first 400 d but then remained nearly constant from 500 to 1000 days. The X-ray flux peaked at about 1030 to 1100 days at $L_x \approx 5.7\times 10^{40}$ \ergs. A power law fit to the X-ray decline after 1000 days gives a power law index of $\alpha = 0.90$.

The light curves of most X-ray supernovae show a decrease with time \citep{dg12, vvd14, drrb16, bocheneketal18} as the supernova shock expands outwards, presumably in a wind medium whose density is decreasing with radius. SN~2014C is one of only a few supernovae that show an  increasing X-ray luminosity with time. Since thermal X-ray emission depends directly on the square of the ambient density, the increasing X-ray emission can be associated with an increasing density in the ambient medium. An increasing density with radius can also be produced in a phase of decreasing mass loss rate, but the rise in X-ray emission would not be as sharp \citep{dg12}. 

There may be several components that contribute to the X-ray emission: the shock in the wind of the progenitor, the forward shock in the dense CSM, or the reverse shock from the interaction with the CSM. The geometry of these components is not necessarily spherical and could be distributed in a more complex way. The shock in the wind of the progenitor, the density of which is expected to decrease with time, would not be expected to give rise to the observed increasing X-ray emission. The forward shock interacting with a high-density medium would be the most straightforward explanation for the rise in X-ray emission. On a longer timescale, the contribution to the X-ray flux from the reverse shock may be expected to dominate at some time after the shock has interacted with the dense CSM, as suggested by simulations, and analysis of the emission line profiles, of SN~1996cr \citep{ddb10, QV19}.

There is distinct evidence in the X-ray data for the onset of interaction of the supernova forward shock with a dense CSM. It is not clear, however, when the shock transmitted into this dense material emerges from this region or even if it does emerge (a strongly radiative shock could be captured in a dense shell). One line of reasoning may be that the forward shock emerges from dense material sometime after 1030 days, when the X-ray luminosity begins to decrease with time. It is possible, however, that before this epoch, the reflected shock from the interaction begins to dominate the X-ray emission, covering the fact that the transmitted shock had emerged much earlier, as was the case in SN 1996cr (\citep{ddb10}). Alternatively, the decrease in X-ray flux may be due to the fact that the density of the region emitting X-rays decreases with time, and the shock has not yet emerged from a high density region. All these factors make it difficult to decide when or if the shock actually emerged from the dense region initially encountered by the forward shock without recourse to simulations and observations at other wavelengths. Thus, from the X-rays alone it is difficult to estimate the thickness and density structure of this high density region. 

If the strong X-ray luminosity is associated with emission from the supernova forward shock, then the deduced temperature can be related to the shock velocity. Table \ref{tab:obslog} shows that the X-ray temperature is initially $> 25$ keV and declines over 2000 days to $\approx 8$ keV. This corresponds to a shock velocity $>5000$ \kms\ declining to ${\sim}3000$ \kms, assuming the density is high enough for the electrons and protons to equilibrate. Otherwise the X-ray temperature gives a lower limit to the velocity. The column depth also declines over this time. The early high column depth coupled with the high temperature suggests a high-velocity shock propagating into a CSM of high density. 

A shock velocity of ${\sim}3000$ \kms\ is reminiscent of the velocity width we determine for the \foiii\ lines and perhaps the helium lines. This in turn suggests that the X-rays arise from the same location as the \foiii\ lines, the reverse shock interacting with the inner ejecta. Perhaps the X-rays arise in the forward shock in the CSM at early times, and from the reverse shock reflected from the dense CSM after 500 - 800 days. X-rays arising at late times in the reverse shock could account for the large iron abundance at later time, but not at early times unless the CSM is contaminated by mixing with the ejecta. Asymmetries may complicate this interpretation. This velocity exceeds the velocity width we determine for \Ha. 

\section{New Radio Observations} \label{sec:radio}

We made a new X-band radio observation of SN~2014C with the Karl G. Jansky Very Large Array (hereafter referred to as the VLA) in the A configuration on 31-Aug-2019 that corresponds to phase $\phi = 2063$~d since first light.  These observations were centered on 9 GHz with a total bandwidth of 2 GHz.  3C48 (J0137$+$331) was utilized as the primary flux calibrator and J2216$+$3518 was used as the secondary or phase calibrator.  The data were processed by the NRAO Pipeline for VLA observations using CASA. We measure a peak flux density of $14.37\pm 0.02$ mJy/beam and the total integrated flux was measured to be $14.81\pm 0.02$ mJy.

The spectral index ($\alpha$, $S_{\nu} \sim \nu^{\alpha}$) of the source has evolved from $\alpha \sim -0.0$ taken near day 1,000 after explosion as reported by \cite{Bietenholz21} to roughly $\alpha \sim -0.6$ near day 2,000 after explosion corresponding to our new observation. This change in spectral slope indicates a synchrotron-emitting source in a relatively ``optically" thin medium. We do not see significant radio absorption at centimeter wavelengths by CSM along the line of sight to the source.

\citet{Bietenholz21} found the average time decay parameter, $\beta$, where $S_{\nu} \propto t^{-\beta}$, to be $\beta \sim 0$ at $\phi \approx 1,000$~d. We determine the value to be $\beta \sim -0.7$ comparing the flux at $\phi \approx 1,000$~d and ours taken roughly 1,000 days later. \citet{Bietenholz21} suggested that SN~2014C was beginning to overrun the densest regions of the CSM at the epoch of their observation. In contrast, the declining X-band radio emission is consistent with a gradual decrease of the density. This may indicate that the density structure of the surrounding medium has changed between 1000 and 2000 days. One inference from the observed decay parameter of \sn\ is that the supernova shock was still interacting with the CSM surrounding SN~2014C at the time of our observation.

\subsection{Constraints from radio observations} \label{subsec:radiocons}

The radio time-decay parameter can be related to the history of mass loss of the progenitor system \citep{Weiler02}. The radio time decay between $\phi \sim 1000$ d and ${\sim}2000$ d, $\beta \sim 0 - -0.7$, is very slow when compared to the (rather sparse) sample of radio observations of Type~IIn supernovae. \citet{Weiler02} found a $\beta$ value of $-1.65$ for Type IIn SN~1986J. \citet{williams02} found that the decay parameter evolved from $-1.22$ to $-2.73$ between 1,000 and 2,000 days after explosion of SN~1988Z. The implication is that the progenitors of SN~1986J and SN~1988Z underwent increased rates of mass loss with time over the last few thousand years before explosion. For Type II SN~1981K, \citet{Weiler02} derived a smaller $\beta \approx -0.70$, comparable to the value that we determined for SN~2014C. Apparently while the mass loss rates for SN~1981K and \sn\ increased with time, they did so less severely than for the two SN~IIn. Note that both the X-ray luminosity discussed in \S \ref{subsec:xconstraint} and the radio luminosity considered here require a decreasing density in the phase $\phi \sim 1000$ d and ${\sim}2000$ d. This does not necessarily mean that the shocks producing radiation in those bands is co-local, but they might be. We also note that the high X-ray luminosity requires a high-density medium, while the radio luminosity does not necessarily. The X-ray flux at this epoch may arise from the reflected shock and the radio from the shock in the outer wind. These factors allow for the possibility that in this epoch the X-rays and radio fluxes arise from different structures.

The radio data hint at some inconsistencies that must be reconciled. The spatially-resolved VLBI data from about 5 years after explosion show a large radius of the shock front, ${\sim}2\times 10^{17}$ cm, and a high velocity, ${\sim}9,400$ \kms\ \citep{Bietenholz21}, that demands expansion into low density material long after the shock collision with a dense CSM produced the first IR, radio, X-ray and then \Ha\ emission. This shock speed is faster than other Type IIn at about the same epoch. \citet{schinzel09} measured a shock speed nearly an order of magnitude slower about two years after optical discovery for the Type Ib/c SN~2001em.

The combined radio observations of SN~2014C thus indicate that the early AMI data and the later VLBI data arise from two spatially separated components, perhaps suggesting departures from spherical symmetry.

\section{Synthesis} \label{sec:syn}

Our multi-year collection of optical data on \sn\ combined with data from other bands raises a number of issues. What is the origin of the broader \Ha\ and why is the associated velocity width of ${\sim}2000$ \kms\ less than that of all the other broadened lines? What determines the line width, ionization state, and temporal evolution of the lines of other elements? How is the velocity width of the \Ha, or any of the other optical lines, reconciled with the expansion velocity implied by the VLBI observations \citep{Bietenholz21}? The large IR luminosity seems to dominate the bolometric luminosity; how is that flux generated? We address some of the relevant issues here and perforce leave others for future investigation.

\subsection{Velocities} \label{subsec:vel}

In \S \ref{subsubsec:profiles} we expressed the widths of various lines in terms of a FWHM. It is, however, unclear how to interpret the FWHM. The CSM structure of \sn\ could be asymmetric, expanding non-homologously, and rife with gradients in composition, temperature, and density. A popular exercise, in which we engaged in \S\ref{subsubsec:profiles}, is to fit emission line profiles with multiple Gaussian components. While it is convenient to fit Gaussians, it is not clear they have anything directly to do with the physics of our problem, and in any case the FWHM might be a measure of a temperature or turbulent velocity, not an expansion velocity \citep{jerk17}. 

\begin{figure*}
    \centering
    \includegraphics[width=\textwidth]{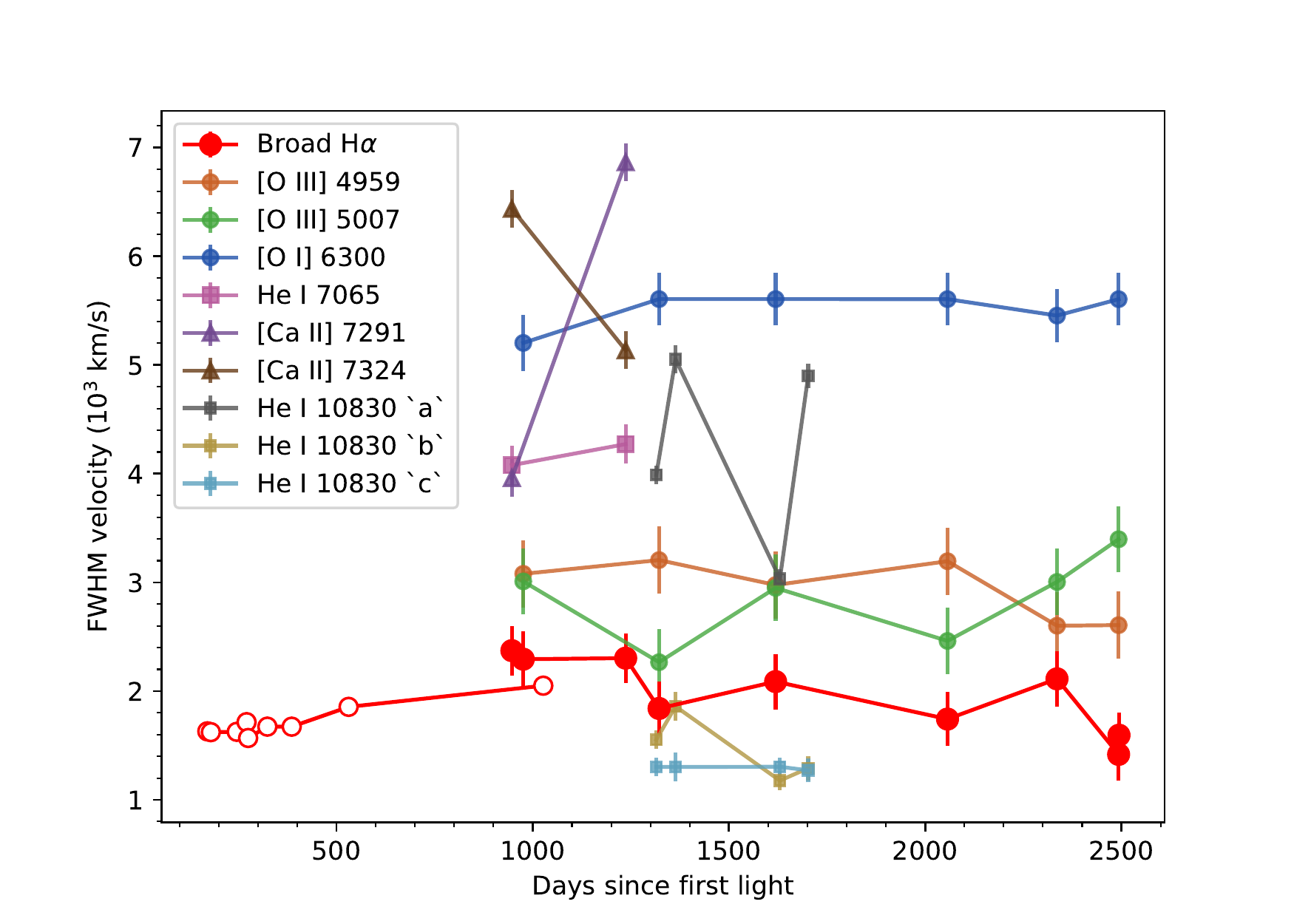}
    \caption{The full-width half maximum velocity evolution as derived from the emission lines that show broadened components in our HET/LRS2 spectra of SN~2014C. The velocities are derived from the FWHM of multi-component Gaussian fits to the observed spectral line profiles. Error bars include the systematic uncertainty from our LRS2 spectral resolution and the statistical error from the MCMC multi-Gaussian fits. The majority of the velocity widths shown here are derived from our new HET spectra. Exceptions include earlier \Ha\ data (open circles) that have been obtained from WISE-REP (\citealt{Milisavljevic15}, \citealt{Anderson17}, and \citealt{mauerhan18}). We also derive three components of the infrared \ion{He}{1} $10830$ \AA\ from data that were obtained by \cite{tinyan19}. We derive a fifth Gaussian component to the \Ha\ emission complex from data obtained by \citet{mauerhan18}. The constant, low velocity of the \Ha\ relative to the other transitions is discussed in the text.
    }
    \label{fig:velocities}
\end{figure*}

Nevertheless, in order to put broadened lines of different wavelength in a commmon perspective, we need to formally convert the FWHM to velocity space. Despite the caveats expressed above, we convert the FWHM of our Gaussian line profile fits to velocity space and qualitatively associate some of those velocities with expansion speeds of the ejecta or post-shock matter.  Hereafter we will refer to the formal velocities associated with the FWHM of an emission line as a {\it velocity width} to underline these ambiguities. Figure \ref{fig:velocities} illustrates the FWHM of the broad components that we determine from multi-component Gaussian fits to various lines in our spectra as detailed in \S \ref{subsubsec:profiles}, but expressed as a velocity width.

As shown in Figure \ref{fig:velocities}, the velocity width of the \Ha\ line may slowly grow to about $\phi = 1000$~d and then gradually shrink, but remains at ${\sim}2000$ \kms\ within one to two $\sigma$, beginning at its first appearance at $\phi = 127$~d until our last observation at $\phi = 2494$~d. This value of the  ``intermediate" width of the broader \Ha\ is often associated with electron-scattering profiles in other supernovae, but we could not firmly establish that any profiles were Lorentzian.

\begin{deluxetable}{r|rrrr}
\label{tab:fwhm_ha5}
\tablecaption{Observed centroid wavelengths and corresponding velocities of the fifth \Ha\ component.
}
\tablehead{\colhead{$\phi$} & \colhead{$\lambda_{\rm obs}$} & \colhead{$\Delta\lambda_{\rm obs}$} & \colhead{$v$} & \colhead{$\Delta v$} \\
\colhead{(days)} & ($\rm{\AA}$) & ($\rm{\AA}$) & (\kms) & (\kms) 
}
\startdata
 275 & 6554 & 0.21 & -420 & 10 \\
 324 & 6555 & 0.21 & -360 & 10 \\
 386 & 6565 & 0.38 &   70 & 20 \\
 531 & 6575 & 0.11 &  540 & 10 \\
\enddata
\end{deluxetable}

As illustrated in Figure \ref{fig:velocities}, the \ion{He}{1} $\lambda$7065 line shows a velocity width of ${\sim}4000$ \kms\ that is similar to that of the main broad component `a', of the \ion{He}{1} 1.083 \mic\ line measured by \citet{tinyan19}. We suspect an additional source of uncertainty that is hidden in the covariance between the FWHM of the `a' and `b' components from the fit. A symptom of this is visible as the co-varying `a' and `b' velocity widths in Figure \ref{fig:velocities}. This degeneracy makes it more difficult to precisely disentangle the true FWHM of the `a' and `b' components. In any case, the helium `a' component velocity width is about twice that of the \Ha\ line. The relatively high velocity width suggests that these He lines arise in a different component from the \Ha\, presumably the ejecta, but further evidence is needed to confirm that supposition.

Figure \ref{fig:velocities} shows a nearly constant velocity width of ${\sim}5500$ \kms\ for the [\ion{O}{1}] $\lambda\lambda 6300, 6364$ doublet. This is nearly twice that of \foiii\ $\lambda 4959$ and $\lambda 5007$, for which we measure ${\sim}3000$ \kms, in agreement with \citet{Milisavljevic15}. Even these slower metal lines exceed that of \Ha\ by a factor of ${\sim}50\%$. With larger scatter, we find that the [\ion{Ca}{2}] lines have a similar velocity width to the [\ion{O}{1}]. All these metal-line velocities might be characteristic of the ejecta, but again there is no firm evidence to make that connection. 

The features of \ion{He}{1} fall midway between the \foiii\ lines and those of [\ion{O}{1}] in velocity space. Since we roughly expect helium to be at larger radii in the ejecta, the somewhat smaller velocity width of helium compared with oxygen and calcium may suggest that the helium has been subject to some deceleration by the CSM. The \ion{He}{1} lines must be non-thermally excited by photoionization, perhaps by radiation from the reverse shock, or by collisional excitation.

\citet{tinyan19} also identified two sub-components, `b' and `c,' of the \ion{He}{1} $\lambda 10830$ line (\S \ref{subsubsec:IR}). The strength of the sub-components relative to the broadest He component is greater at $\phi = 1368$~d than at $\phi = 1707$~d. The velocity widths of the sub-components of \ion{He}{1} are ${\sim}1500$ \kms\ for blue-shifted (-4000 \kms) component `b' and ${\sim}1200$ \kms\ for rest component `c.' The velocity widths of components `b' and `c' are roughly half that of \Ha\ (Figure \ref{fig:velocities}).  \citet{tinyan19} argue that these components are from shocked CSM, with the component at -4000 \kms\ related to the VLBI hotspot identified by \citet{Bietenholz18}. \citet{Bietenholz21} argue, however, that evidence for a hot spot, or any asymmetry in the VLBI image, may be an artifact of the observation/reduction process; there is still some East/West asymmetry.  

Upon inspecting the broadened \Ha\ profile at dates that precede our HET observations, we identified an anomalous fifth component to the emission complex (see \S \ref{subsubsec:profiles}), the central wavelength of which appeared to shift to the red, across \Ha, between $\phi = 275 - 623$~d. In order to identify whether this might be a third \Ha\ sub-component emitted from material with some peculiar bulk velocity, we included a fifth component to our Gaussian models at those pertinent epochs (Figure \ref{fig:benfit}, bottom panel). We derive a velocity from the relative centroid shift of the fifth sub-component of -406 \kms\ at $\phi=275$~d, which increases monotonically, with some small deviations from linearity, until it reaches +540 \kms\ at $\phi=531$~d, after which it apparently disappears. There is also some weak evidence of a fifth component to the red of the \Ha\ complex in our $\phi =2493$~d and $\phi=2494$~d spectra. The interpretation of this ``moving" sub-component is unclear. We present the derived centroids of the sub-component and corresponding velocity shifts in Table \ref{tab:fwhm_ha5}.

The substructure in the decomposition of \Ha\ is not connected in any direct way with that of the substructure of \ion{He}{1} $\lambda 10830$. In \Ha, the sub-components are separated from the rest wavelength by about 400 \kms, compared to sub-component `b' of \ion{He}{1}\ $\lambda 10830$ with displacement about 4000 \kms. \cite{tinyan19} found evidence of sub-component `b' to the \ion{He}{1} line at phase $\phi = 1315-1702$~d. The fifth travelling sub-component to the \Ha\ emission that is evident in public optical spectra shows up early relative to the \cite{tinyan19} NIR observations that exhibit component `b' such that the two are not contemporaneous. Our HET/LRS2-B spectra that are contemporaneous with the \cite{tinyan19} observations do not show significant evidence for a fifth component to the \Ha/[N II] emission complex. This may indicate that the two phenomena are of separate physical origin. Given the different phases, velocities, and velocity uncertainties of these extra sub-components, it seems likely that these anomalous emissions in \Ha\ and \ion{He}{1} may originate from different sources.

We find velocity widths $\gtrsim 3000$ \kms\ for all the major broad lines in the optical and NIR except \Ha, which in contrast shows a relatively low velocity width of ${\sim}2000$ \kms\ across the duration. We interpret this as evidence that the \ion{He}{1} $\lambda 7065$, \ion{He}{1} $\lambda 10830$, \foiii, [\ion{O}{1}], and [\ion{Ca}{2}], are emitted from the hydrogen-deficient inner ejecta that is excited by the inward-travelling reflected shock after the forward shock has collided with the CSM, while the \Ha\ is emitted elsewhere in the CSM. A caveat to this interpretation is that for a spherical reverse shock the material interior to the reverse shock should be expanding homologously with $v \propto r$. As the reverse shock propagates inward in mass, the metal lines from the ejecta irradiated by the reverse shock should slow and narrow with time. This assumes that only material close to the reverse shock is radiating, but that depends on the optical depth of the ejecta. In any case, we see no sign of such an evolution in the width of the metal lines. 

We also note that none of the velocity widths portrayed in Figure \ref{fig:velocities} are comparable to the high velocities $> 9,000$ \kms\ determined directly by the VLBI observations of \citet{Bietenholz18} and \citet{Bietenholz21}. The radio emission apparently comes from a completely different region than the optical emission lines. This is difficult to reconcile with a spherically-symmetric model. 

While the physical meaning of the FWHM of our features remains ambiguous, the full width at the base of a broad feature may place some constraint on the maximum shock velocity. As noted by \citet{Milisavljevic15}, the base of the \Ha\ line at $\phi = 386$~d extended from -2000 to +2200 \kms, thus setting limits on the velocity of the forward shock in the hydrogen-rich CSM.

The lines of the metals most plausibly arise in the ejecta and are most probably excited by the hard flux from the reverse shock that results from collision of the ejecta with the dense CSM.

\subsection{Light Curves} \label{subsec:lc}

\begin{figure}
    \centering
    \includegraphics[width=\columnwidth]{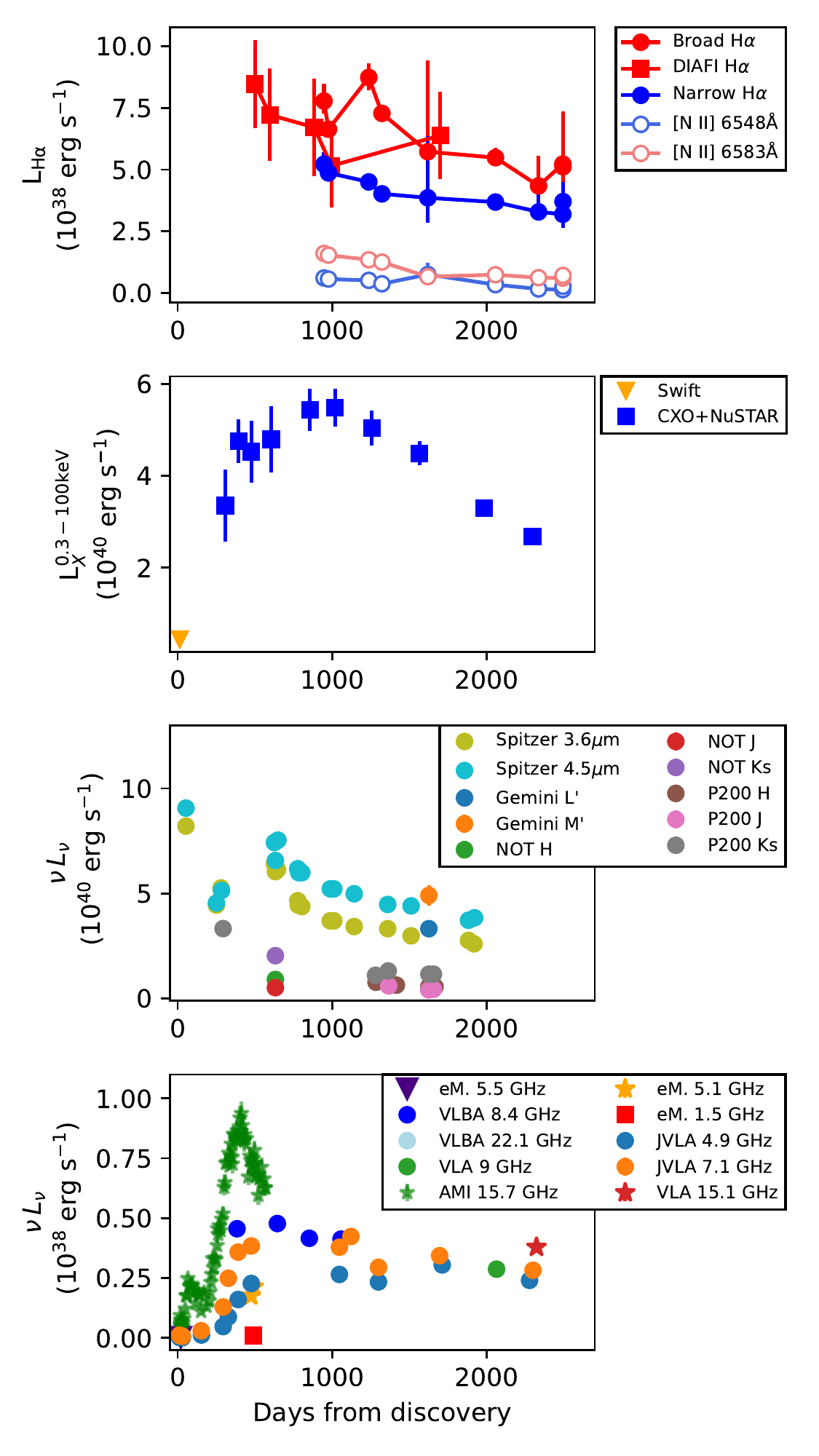}
    \caption{Light curves of SN 2014C at optical (this work), X-ray (this work), infrared \citep{tinyan19}, and radio (this work and \citealt{Anderson17}, \citealt{Bietenholz18}) wavelengths, in that order from top to bottom. The optical luminosities are derived from multi-component Gaussian fits to the emission complex around \Ha\ for the points shown as filled circles, while filled squares are derived from our narrow band images obtained with the DIAFI instrument mounted on the 2.7m HJS telescope. There is some weak evidence of a decline in the broad \Ha\ spectra, and narrow lines also tend to decrease in luminosity. The uncertainties on the optical points are propagated from the uncertainty from our LRS2 flux calibrations. When coupled with the apparent decline in the independently-derived DIAFI photometry (which is also at low confidence) we interpret this decline as real.}
    \label{fig:lightcurves}
\end{figure}

The top panel of Figure \ref{fig:lightcurves} shows the \Ha\ light curve, spectroscopically-derived luminosities from our LRS2 spectra. The red squares are based on our DIAFI narrow-band photometry that has accurate calibration to flux standards. The red circles are derived from our HET spectra. The calibration of the latter points is somewhat more uncertain, but the two sets of data are substantially consistent. The \Ha\ light curve of the broader line may show some evidence of a decline, although this is weak given the large uncertainties from the flux normalisation. We find a similar low confidence decline in the independently derived DIAFI data. In combination with similar declines seen at other energies, we interpret the \Ha\ decline as modest but real. 

Figure \ref{fig:lightcurves} also shows the narrow line luminosities of H$\alpha$ and the \fnii\ $\lambda \lambda$6548, 6583 emissions for comparison. These luminosities were derived by first computing the integrated fluxes of the Gaussian distribution fits to those lines. We then transform those fluxes to a luminosity given the luminosity distance and redshift of the source. The spectra from which these luminosities are derived are corrected for Milky Way extinction, but we make no correction for extinction from the host galaxy. Given the low redshift of the source, we also assume that the K-correction is negligible.

We also present comparable luminosities at X-ray, infrared and radio wavelengths, computed by us as well as taken from the literature \citep{Milisavljevic15, Margutti17, Anderson17, Bietenholz18, mauerhan18, Bietenholz21}. 

Figure \ref{fig:lightcurves} shows that the X-rays and mid-IR dominate the bolometric luminosity. The 4.5 \mic\ band luminosity may slightly exceed the X-ray luminosity around $\phi = 600$~d, the two are roughly comparable at $\phi = 1000$~d, and the IR luminosity again slightly exceeds the X-ray at $\phi = 2,500$~d. Between $\phi = 1000 - 2000$~d, the \Ha\ is less than the IR and X-ray luminosity by about two orders of magnitude and the radio by yet another order of magnitude.

The origin of the strong IR luminosity, presumably by heating of dust, is not completely clear. The data of \citet{tinyan19} show a dip at $\phi = 250$~d corresponding to peak dust temperature and at about the same time as the early dip in the radio and the onset of the X-rays. The IR luminosity then shows a higher flux at about $\phi = 600$~d that corresponds to no peak feature in data at other wavelengths. This epoch roughly corresponds to when the X-ray light curve shows a brief flattening and when the 15.7 GHz radio flux may halt its steep decline. Because of a gap in the data, a peak in the IR data coinciding with the peak of the 15.7 GHz data at ${\sim}400$ d cannot be ruled out. Some of this temporal behavior may result from noise in the respective bands.

While the origin of the radiation in the various bands is likely to involve different locations and different physics, we attempted a comparison of the rate of decline at later times by performing a linear fit to the luminosity in the different bands illustrated in Figure \ref{fig:lightcurves} in log-log space to derive the power law index of each of the declines. We find a rapid decline in the late-time X-ray light curve (power-law index $\alpha = 0.90$ at $\phi > 1000$~d) that contrasts with the slower decline of the \Ha\ ($\alpha = 0.36$ at $\phi \geq 947$~d) and radio ($\alpha = 0.38$ $\phi > 1000$~d) light curves, while the IR light curve favours an intermediate value ($\alpha = 0.51$ at $\phi > 765$~d).

We integrate the X-ray and infrared luminosity curves to approximate and compare the total energy emitted at these different regions of the SED. For the infrared we use only the well-sampled Spitzer 3.6 \mic\ and 4.5 \mic\ bands, deeming other bands to contribute a subdominant proportion of the luminosity. We compute the total energy emitted in the synthetic X-ray band 0.3-100 keV to be $9.35\times10^{43}$ erg between $\phi=307-2297$~d. We find a total energy emitted in the combined Spitzer 3.6 and 4.5 \mic\ bands of $18.05\times10^{43}$ erg between $\phi=53-1922$~d. The total emitted IR energy is essentially double that emitted in the X-ray, despite the slight temporal offset between these measurements. We note that the emitted IR energy we have estimated here is a lower bound as we have omitted bands other than the Spitzer 3.6 \mic\ and 4.5 \mic\ bands. If those bands were included, the total emitted IR energy would dominate even more over the emitted energy at X-ray and other wavelengths. We have opted not to fit, for example, a modified black-body model here as only two bands are available at the majority of epochs. This would lead to overfitting with a black-body model of two parameters (the radius and temperature).

\citet{harris20} noted that the \Ha\ emission was detected prior to the rise in the radio at 186 d. They proposed that the rise in radio flux occurred after the forward shock had departed a dense shell and was propagating in the outer CSM. That hypothesis seems difficult to reconcile with the similar epoch of onset and continued high luminosity of X-rays. 

\subsection{Common Envelope Ejection and a Toroidal CSM} \label{subsec:cetorus}

The central conundrum revealed by our extensive observations of the optical spectra is the nearly constant value of the FWHM of the \Ha\ line with a velocity width of ${\sim}2000$~\kms\ that is not shared by any of the other prominent optical/IR emission lines nor by the expansion directly measured by VBRI at similar epochs. The radio expansion velocity is $v = 13,040 \pm 690$ \kms\ at 1000 days and $9,400 \pm 2,900$ \kms\ at 1700 d \citep{Bietenholz21}. If the \Ha\ velocity width is related to a shock velocity, this is a strong hint that the CSM of \sn\ has a complicated, non-spherical geometry. There is clearly a dense, hydrogen-rich CSM, but whether there is a distinct spherical shell is far less clear.

Different techniques result in different estimates of the density structure with distributions ranging from constant to declining as $\rho \propto r^{-3}$ \citep{Margutti17, harris20, tinyan19, Brethauer20, Bietenholz21, vargas21}. Whatever the origin and morphology of the CSM, it can only have one density profile if it is spherically-symmetric. The disagreement among the various estimates of the density profile does not establish that the CSM departs from spherical symmetry, but leaves open the possibility of substantial morphological asymmetry with various wavelength ranges sampling different density distributions. Another implication is that caution should be exercised in taking  any of the density distributions cited in the literature literally, including a thin, dense shell. At the same time, the spatially-resolved VLBI observations of \citet{Bietenholz21} suggest that the locus of the shock producing that radio flux is substantially spherical (or at least circularly symmetric).

The detection of the strong broad \Ha\ at $\phi = 127$~d shows that the interaction with some hydrogen-rich material was already underway at that time. Sparse temporal sampling, different production mechanisms, and different sensitivities in the optical, radio, and X-ray bands makes it difficult to tell from the data when the collision with the CSM occurred.

\begin{figure*}
\centering
\includegraphics[width=0.7\textwidth]{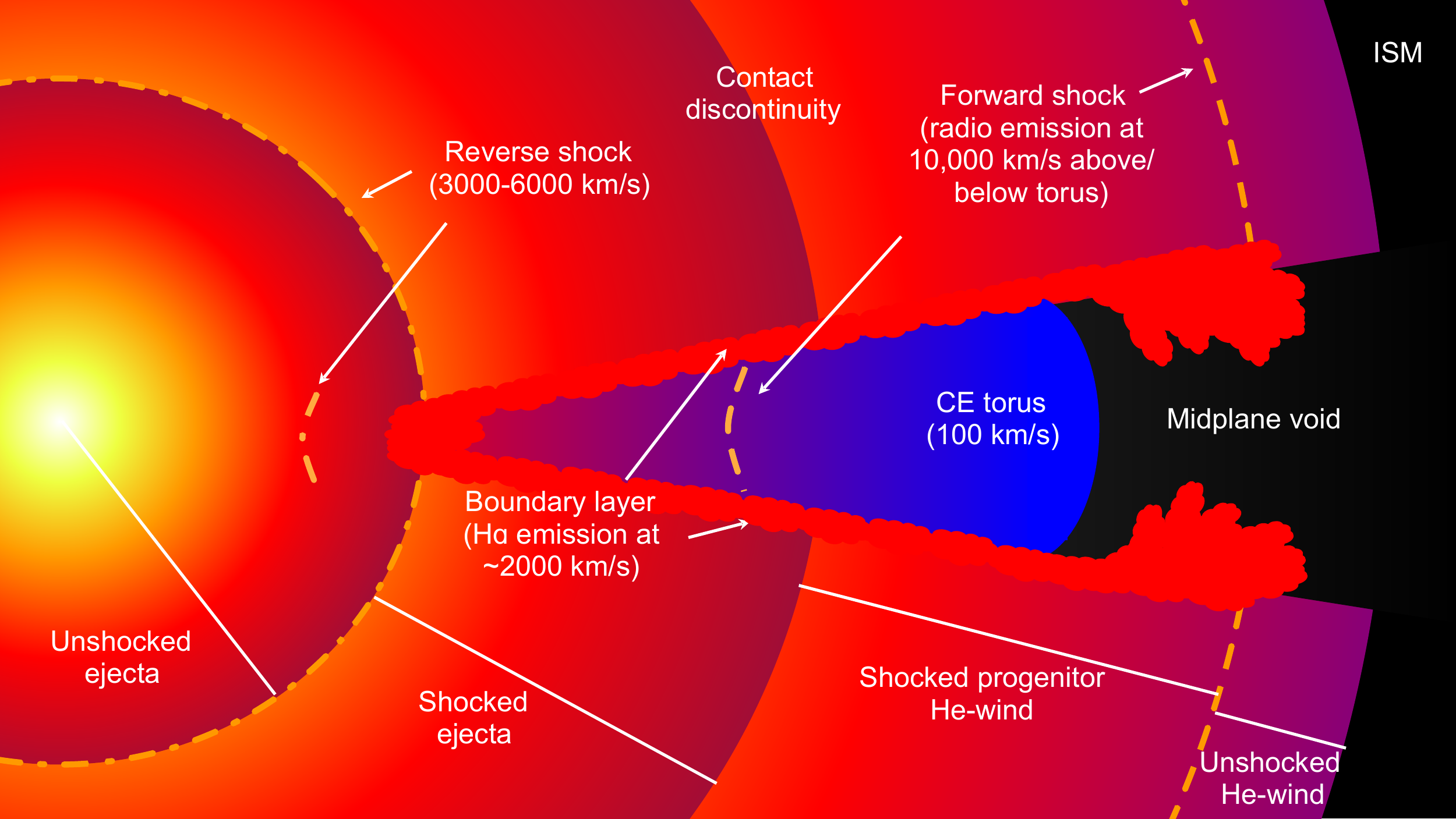}
\caption{Schematic of our proposed geometry of SN~2014C. We suggest that the \Ha\ emission originates from a boundary layer between the common-envelope torus and the shocked ejecta/He-wind from the progenitor. This may reconcile the relatively slow \Ha\ velocity width of ${\sim}2000$ \kms\ that we measure with the faster emission lines ([\ion{O}{1}], [\ion{O}{3}], \ion{He}{1}, [\ion{Ca}{2}]), that we designate to the reverse shock receding back into the ejecta. The radio velocity from \cite{Bietenholz21} of ${\sim}$10,000 \kms\ corresponds to the quasi-spherical forward shock propagating in the progenitor He-wind. The boundary layer between the torus and shocked ejecta/He-wind is subject to Kelvin-Helmotz instabilities, the inner edge of the torus is subject to Richtymer-Meshkov instabilities, and the contact discontinuity is subject to Rayleigh-Taylor instabilities (not shown). The putative secondary star is also not shown. The viewing angle favored by the observations may be at about 60 degrees from the pole (\S \ref{sub:originIR}). }
\label{fig:schematic}
\end{figure*}

Given various inconsistencies in the multi-wavelength data in the paradigm of a spherically-symmetric CSM, we need to consider possible asymmetric distributions. The hydrogen deficiency and rate of explosions of stripped envelope supernovae suggest that they arise in binary evolution \citep{Li2011, snex}. The fact that \sn\ was originally of spectroscopic Type Ib thus points to a role for binary evolution, a possibility discussed by \citet{Margutti17}. \citet{tinyan19} noted that in the first 800 days the evolution of the inferred dust mass was consistent with pre-existing CSM dust heated radiatively or collisionally by the shock interaction with a CSM shell of constant density. They proposed that the rapid expansion of the shock indicated by the VLBI observations of \citet{Bietenholz18} could be the result of an anisotropic CSM that allowed parts of the forward shock to propagate freely and discussed binary evolution as the source of that anisotropy.

While some asymmetries may be produced by single stars, we will thus examine a scenario in which binary evolution led to a common envelope phase that was responsible for the loss of the hydrogen envelope \citep{sunmc20} and formation of the hydrogen-rich CSM. The likely distribution of matter in a system that has undergone binary evolution with the ejection of a common envelope is that the hydrogen-rich envelope material substantially will be confined  to the equatorial plane. The geometry of the CSM may be that of a fat torus \citep{lawsmith20}. 

We consider a hypothetical toroidal geometry of the progenitor system, a schematic of which is provided in Figure \ref{fig:schematic}. Similar models have been discussed by \citet{Smith11iqb} and simulated by \citet{suzuki19}. In this picture, the helium star supernova progenitor blew a fast wind that interacted with the main sequence secondary that facilitated the past expulsion of the progenitor's hydrogen envelope in a common envelope interaction. The secondary blows a slower hydrogen-rich wind that would be entrained by the fast hydrogen-poor wind of the primary, thus forming a bow shock and a tail. The secondary wind tail prior to explosion would probably be an open spiral in the centre-of-mass rest frame.

The inner edge of the expelled progenitor envelope would have a dense ring created by the interaction of the progenitor wind with the dense envelope material. Beyond that interaction region, the toroidal envelope would expand homologously at the escape velocity from the binary system ${\sim}100$ \kms\ \citep{lawsmith20}. At higher latitudes, the fast progenitor wind would continue to flow in a quasi-spherical fashion. The fast wind would connect to the toroidal material through a boundary layer that may engender various fluid instabilities. 

After the explosion, the progenitor helium star would have formed a pulsar or magnetar; a pulsar wind nebula could contribute to the ionization and excitation structure of the CSM \citep{chevfran92,Milisavljevic12au}. A relatively massive main sequence secondary star is likely to remain nearby or even bound after the explosion with its wind now being ablated and swept up by the ejecta.

In the proposed dusty torus CSM structure, the forward shock will proceed more rapidly at higher latitudes and will be decelerated most severely in the equatorial plane. The toroidal geometry allows room in the polar direction for the continued expansion of the ejecta in the low-density, hydrogen-deficient wind of the progenitor. The reverse shock will also have a complex geometry that could be far from spherical, with small radius in the equatorial plane, but extending further in more polar directions. A contact discontinuity with a similar distorted shape would fall between the forward and reverse shocks. X-rays could be coming from both the forward shock and the reverse shock, neither of which would be expected to have spherical loci. The radio emission resolved by \citet{Bietenholz21} could have a large radius and a quasi-spherical locus because the forward shock is propagating broadly in the wind above and below the equatorial torus. Other radio emission could be coming from the denser gas in the equatorial plane. The forward shock could be subject to Richtmyer-Meshkov, Rayleigh-Taylor, and Vishniac \citep{Vishniac87} instabilities in the midplane and Kelvin-Helmholz and Rayleigh-Taylor instabilities where the ejecta shear along the surface of the torus. 

The recombination time per particle, $t \sim 10^5/n_e$ y, is short for the dense torus we propose in the equatorial plane with densities $> 10^5$ cm$^{-3}$. The short recombination time means this matter has to be continually exposed to photoionizing radiation to produce \Ha\ over the seven years of our observations. The \Ha\ could, in principle, be powered by photoionization from the reverse shock, the forward shock in the equatorial torus, by shocked clumps in the torus, by a pulsar, or by the secondary star. The progenitor helium star and flux from the supernova could also contribute with recombination times of order a year. The photoionizing flux depends on the temperature, density, and composition of the material all of which vary in the geometry we envisage here \citep{chevfran94}. UV flux would be a more effective means of ionization, but estimating that is beyond the scope of this paper.

\subsection{Origin of the Hydrogen Emission} \label{subsec:horigin}

In the CE/torus paradigm, the hydrogen will primarily be confined to the equatorial torus. The supernova shock will expand within the wind of the progenitor star until it impacts the dense torus. The dense CSM torus material is expected substantially to slow the forward shock propagating in the equatorial plane.

\subsubsection{Hydrogen emission from the forward shock} \label{subsubsec:FS}

A sufficiently dense equatorial CSM is capable of decelerating the forward shock to the level observed for the \Ha. Some of the \Ha\ emission thus could come from behind the decelerated forward shock as it propagates into the midplane of the torus.  There are, however, several issues with the suggestion that this be the source of the observed \Ha\ emission. A principal problem is that the midplane portion of the shock should continue to decelerate. This conflicts with the nearly constant velocity width we observe. Lines from a recently shocked region also should all show about the same velocity, whereas we observe \Ha\ to have an appreciably lower velocity than other broad lines. Any new ``intermediate'' ${\sim}2000$ \kms\ component from metal lines in the recently-shocked outer CSM could be hidden under the ``broad'' ${\sim}3500$ \kms\ component from the reflected shock, but this remains to be established. 

\subsubsection{Hydrogen emission from the companion} \label{subsubsec:hacomp}

Any secondary star will survive the explosion either still bound to the compact remnant or unbound but nearby. \citet{sunmc20} computed binary evolution models matching the lifetime of the host star cluster and susceptible to common envelope evolution. Two models had initial secondary mass of 2 - 3 \msun\ with final secondary masses of ${\sim}1.8$ and ${\sim}4.6$ \msun. The final separation was $2 - 3\times10^{13}$ cm, about half of the initial separation. The final orbital periods were about 300~d. One was nearly unbound, the other was probably still bound. The final separations were sufficiently large that the effect of impact heating of the companion is expected to be negligible \citep{WLM75,ogata21}. The companion is thus expected to retain its ZAMS luminosity. From the models, the companion will be about 30th magnitude, too dim to easily detect. Typical orbital velocities if still bound, ${\sim}10$ \kms, are too small to be directly related to the motion of the fifth \Ha\ Gaussian component or the substructure of the He I $\lambda$10830 line that are of order several 100 to 1000 \kms (\S \ref{subsec:vel}). The length scale of the orbit is also too small to be related to the CSM density perturbation length determined by \citet{vargas21} to be ${\sim}10^{16}$ cm. 

At $\sim10$ \kms, the companion would have moved only ${\sim}2\times10^{14}$ cm in the 7 years since explosion, so would appear essentially to be an unmoving source of \Ha. Simulations of ejecta/companion interactions for conditions relevant to \sn\ suggest that less than $10^{-2}$ \msun\ will be ejected from a companion of ${\sim}10$ \msun\ \citep{Hirai18}. In addition, the ablated material will have a velocity of $\lta 1000$ \kms\ and perhaps asymptotically as low as ${\sim}10$ \kms\ (R. Hirai, private communication, 2022). This suggests that while hydrogen stripped from the companion might contribute to the narrow \Ha\ feature, it is unlikely to contribute to the broader feature with FWHM $\sim 2000$ \kms\ that we prominently observe.

\subsubsection{Hydrogen emission from the boundary layer} \label{subsubsec:BL}

Another source of the \Ha\ emission is the boundary layer between the ejecta and the torus that blankets both surfaces of the torus. 

\citet{suzuki19} presented a 2D radiation dynamical model of a supernova exploding into an equatorial torus. This model is not directly applicable to \sn\ because the model torus is compact, with an outer radius of just $5\times10^{15}$ cm, but the torus mass is comparable, a few \msun, and the opening angle of 10 - 20 degrees is possibly relevant. The radiative transfer is somewhat simplified and ignores dust, but some characteristics of the models, aspect angle effects and line profiles, may be applicable qualitatively to \sn. 

As expected, in the models of \citet{suzuki19} the forward shock propagates nearly spherically in polar directions and is inhibited in the equatorial plane. A ``void" forms in the equatorial plane beyond the outer edge of the torus with an opening angle that slightly exceeds that of the torus. Near the ejecta/torus boundary, the ejecta do not expand ballistically; rather, the dynamic interaction of the ejecta and torus affect the dynamics of both the ejecta and the torus material. The details will depend on the vertical structure of the torus that is largely unknown but perhaps illuminated by simulations such as those of \citet{lawsmith20}. The ejecta/torus boundary is subject to the instabilities we outlined in \S \ref{subsec:cetorus} that \citet{suzuki19} argue could contribute to irregularities in the light curve that are more distinct for larger disk masses. It would be interesting if the radial length scales of the Kelvin-Helmholz instabilities were comparable to those deduced by \citet{vargas21}.

Of special importance to our observations, \citet{suzuki19} predict that the most intense flux arises at the boundary layer between the nearly static torus and the rapidly expanding ejecta. Unlike the locus of the forward shock, the boundary layer will be a quasi-time independent structure, as the source of the \Ha\ emission in \sn\ seems to be. The boundary layer could also contribute to IR, radio, and X-ray flux. 

In the simulations of \citet{suzuki19}, the velocity in the boundary layer is greater than the velocity width of the \Ha, but conditions might be different in \sn\ with a more extended torus. The velocity drops rapidly toward the midplane so there will surely be some hydrogen with a speed of ${\sim}2000$~\kms\ somewhere between the boundary layer and the midplane. The question of the density of that layer and its exposure to ionizing radiation will require a deeper study. 

The simulations also suggest another possibility: the void left near the midplane where the ejecta blast past the outer rim of the torus. That region is partially filled with material of substantially lower velocity that could be of order 2000~\kms. The issue would again be the density of any hydrogen there and its exposure to ionizing radiation. This structure would also be quasi-stationary in a manner consistent with our observations of \Ha. 

While it is difficult to put quantitative limits on this possibility, we propose that radiation from the boundary layer is a plausible source of the \Ha\ we observe.

\subsection{Constraints from IR emission} \label{sub:originIR}

An important question is whether the IR observations can usefully constrain or account for the toroidal geometry we have hypothesized. The origin of the IR emission presented by \citet{tinyan16,tinyan19} is especially important because the IR emission appears to dominate the bolometric luminosity. As shown in Figure \ref{fig:lightcurves}, the IR luminosity exceeds the X-ray luminosity at essentially all epochs where they are measured contemporaneously. While X-ray emission can contribute to heating of the dust, the X-ray flux thus apparently cannot account for the majority of the dust emission. The fact that models suggest that the torus/ejecta boundary layer is the source of the most intense flux leads us to look there for an explanation of the dominant source of bolometric luminosity in the IR.

To understand the role of the torus in shaping the observational properties of \sn, it is important to know whether the torus is optically thick. This requires knowledge of the size of the torus and the nature of its opacity.

There is no direct evidence of the outer radius of the equatorial torus we propose for \sn. There are constraints on the location of the sources of emission. \citet{tinyan19} find the black body radius of the dust emission to be ${\sim}1.7\times10^{17}$ cm at $\phi \sim 1620$ d. The torus is presumably larger than that. \citet{Bietenholz21} find a radio shock velocity to be 9,400 \kms\ at $\phi = 1700$ d. By the epoch of our last observation at $\phi = 2494$ d, this would correspond to a position of the shock of ${\sim}2.0\times10^{17}$ cm. The agreement of these radii could suggest some correlated radio and dust emission, perhaps along the ejecta/torus interface. If the torus formed in a common envelope event, it could have a radial velocity of ${\sim}100$ \kms, suggesting that the CSM radiating at ${\sim}2\times10^{17}$ cm was expelled about 500 years ago. 

The optical depth of the gas in the equatorial plane would be of order 
\begin{equation}
    \tau_{gas} \sim \kappa_{gas} \rho_{gas} R \sim 0.1 \kappa_{gas} n_{e,6} R_{17}
\end{equation}
where $n_{e,6}$ is a characteristic electron density in the torus in units of 10$^6$ cm$^{-3}$ (and we have taken $\rho_{gas} = 10^{-24}~n_e$) and $R_{17}$ is the outer radius of the torus in units of $10^{17}$ cm. Even a fully-ionized gas with $\kappa_{gas} \sim 0.2$ cm$^{-2}$ g$^{-1}$ would be optically thin. The CSM is, however, full of dust for which
\begin{equation} \label{eqn:taudust}
    \tau_{dust} \sim \kappa_{dust} \rho_{dust} R \sim 4 n_{e,6} R_{17}
\end{equation}
where we have taken a typical dust opacity to be 4000 cm$^2$ g$^{-1}$ \citep{Draine03, Shirley11} and the dust density to be 0.01 of the gas density. This opacity suggests that the torus could be opaque in the equatorial plane but optically thin in the vertical direction if the thickness of the torus is substantially less than its radius. 

As noted in \S \ref{subsec:spectra}, the appearance of standard optical emission lines from core-collapse ejecta in our data suggests that the environment is optically thin along the line of sight. The line of sight is thus probably not in the midplane of the torus.

Dust in the torus might be heated by the forward shock propagating into the torus, but that process may be inhibited if the disk is optically thick to dust opacity in the radial direction. A torus that is optically thin to dust in the vertical direction would promote the heating of the dust from radiation generated in the boundary layer.

\citet{suzuki19} argue that if the CSM torus is optically thick in the equatorial plane, as suggested by Equation \ref{eqn:taudust}, the bolometric light curve will be sensitive to the aspect angle. A small aspect angle, pole-on, will enable a direct view of both the ejecta and the CSM interaction region and yield a relatively rapid rise and decline in the light curve. An aspect angle near the equatorial plane, 90\degree, will yield a slower rise and decline controlled by the diffusion through the torus plane. A slow rise and decline is also promoted by a more massive and fatter torus. Observations presented in Figure \ref{fig:lightcurves} qualify as a ``slow" decline, only a factor of order 2 in 1500 days. The ``fast" light curves of \citet{suzuki19} decline by an order of magnitude or more over the same relative timescale (several rise times). The IR light curve suggests that \sn\ is interacting with a relatively massive CSM torus of appreciable opening angle, closer to 20\degree\ than to 10\degree, and viewed from an aspect angle exceeding ${\sim}60$\degree. Higher aspect angle also tends to yield lower luminosities. At later times, the disk will become more optically thin thus muting aspect angle effects. 

\section{Conclusions} \label{sec:concl}

We derived spectroscopic information, especially line-width velocities, for all emission lines that display a broadened component to their overarching profile as deduced from our new set of HET/LRS2 optical spectra covering $\phi = 947 - 2494$~d. The velocities were computed using multi-component Gaussian fits, with a Gaussian order chosen by inspection of the observed spectroscopic line profiles. We fit broadened components to the lines of [\ion{O}{3}] $\lambda \lambda$4959, 5007, [\ion{O}{1}] $\lambda$6300, \Ha, \ion{He}{1} $\lambda$7065 and [\ion{Ca}{2}] $\lambda \lambda$7291, 7324 and thereby derived line-velocity information across seven years and throughout the optical spectrum. We also fit the \HeI\ 1.0830 \mic\ line from \citet{tinyan19}.

We derived luminosity information across the same seven years from radio to X-ray, with new measurements at optical and radio wavelengths. This is also the first time the full set of X-ray measurements have been published, using our reduction procedures and analysis steps to arrive at the full X-ray light curve. We also include the full set of infrared spectroscopic observations from \cite{tinyan19}. We took previously published radio and optical fluxes from \cite{Milisavljevic15,  Anderson17, Bietenholz18, mauerhan18, Bietenholz21} and, by careful consideration of the band-widths of the various observations (which are different by orders of magnitude from radio to X-ray) we transformed these fluxes into an equivalent luminosity space of erg s$^{-1}$ to compare the global light curve behaviour of SN~2014C across the majority of the electromagnetic spectrum.

This study has determined a number of factors that give important insights into the physical structure of \sn: 

\begin{enumerate}
    \item The broadened \Ha\ emission profile has a constant velocity width of ${\sim}2000$ \kms\ across the seven years of optical spectroscopic observations that are available both in the previous literature and presented in this study. We have extended the coverage of the \Ha\ emission by an additional 4.25 years.
    \item All other broadened lines we measure show velocity widths larger than \Ha. We find the velocity widths of \foiii\ $\lambda 4959$ and $\lambda 5007$ to be ${\sim}3000$ \kms, \ion{He}{1} $\lambda$7065 and \ion{He}{1} $\lambda$10830 to be ${\sim}4000$ \kms, and the [\ion{O}{1}] $\lambda \lambda$6300, 6364 doublet and [\ion{Ca}{2}] $\lambda$7291 and $\lambda$7324 to be ${\sim}6000$ \kms. 
    \item Observation of emission of metal lines commonly associated with the ejecta of core-collapse supernovae in the first 1000 days suggest the line of sight to the ejecta is optically thin.
    \item The \Ha\ profiles do not show the expected double peak and hence are inconsistent with a simple thin shell model for the \Ha\ emission although such peaks might be lost in the noise.
    \item The broad \Ha\ is centered at zero velocity and hence shows no evidence of dust extinction local to the supernova geometry. 
    \item The luminosity of the broadened \Ha\ component declines slowly for five years, from $\phi = 500-2494$~d post-explosion as suggested by the spectral line flux and confirmed by our flux-calibrated narrow-band imaging.
    \item Both broad and narrow components of the \ion{He}{1} 1.083 \mic\ line are displaced to the red by ${\sim}400$~\kms. This displacement is the opposite of that expected for dust obscuration and in contrast to the lack of any such displacement of \Ha.
    \item \Ha\ and \ion{He}{1} 1.083 \mic\ show atypical sub-components in their line profiles that are apparently unrelated. \Ha\ shows a ``travelling fifth component" at some phases. Component `b' of the \ion{He}{1} 1.083 \mic\ line is displaced to the blue by 4076 \kms.
    \item The narrow [S II] doublet shows a decrease in flux at nearly constant density, suggesting an origin in an H II region hidden within the glare of the supernova image.
    \item The evolution of the luminosities of the radio, infrared, and X-ray emission are roughly consistent with one another, in that they rise up to about $\phi = $ 500, 700, and 1000 days in the radio, infrared, and X-ray, respectively, and then decline throughout the rest of the available epochs up to day $\phi \sim 2400$.
    \item The IR flux seems to dominate the bolometric luminosity. 
    \item Velocities derived from the X-ray shock temperatures are similar to those of some of the metal lines, suggesting that they both arise from the same component, which we equate with the shocked ejecta.
    \item The optical emission lines have much lower velocity widths than that derived from the VLBI radio emission ($> 9000$~\kms), which shows a roughly circularly-symmetric shock front \citep{Bietenholz21}.
    \end{enumerate}

Our extended monitoring of the optical spectrum showing a low, nearly constant velocity width of the \Ha\ emission that contrasts strongly with the high shock velocity determined by VLBI radio observations shows that the CSM is unlikely to be spherically symmetric. In particular, we find that the assumption of a dense spherically-symmetric shell of hydrogen is not consistent with all the data. 

While much more quantitative analysis is required, we propose a multi-component, non-spherical configuration of \sn\ and its immediate circumstellar environment that appears to accommodate the available data. In this picture, the progenitor binary system first expels a hydrogen-rich toroidal common envelope and then a fast, helium-rich wind from the supernova progenitor star. The supernova ejecta then collide with this complex environment. The early X-ray and radio flux arise when the forward shock impacts the inner portions of the CSM torus. The later X-ray flux may arise from the reverse shock that propagates into the ejecta. The later VLBI radio reveals a nearly circular geometry as the forward shock propagates into the quasi-spherical fast wind in which the CSM torus is embedded. We propose that the \Ha\ emission arises in the boundary layers where the ejecta interact with the two surfaces of the torus. The boundary layers are also the likely source of the heating of dust in the torus, the luminosity of which dominates the bolometric luminosity. A surviving companion star may contribute to the narrow \Ha\ emission, and a pulsar may contribute to some of the emission lines of high ionization. Such an environment for the production of radio, infrared, optical and X-ray flux is much richer and more complex than previously considered for \sn.

To properly explore the interaction of the explosion of \sn\ with a companion star and a CSM concentrated in the equatorial plane and to account for the multi-wavelength spectra requires a multi-dimensional radiation hydrodynamic calculation that is beyond the scope of the current paper. 

Future observations of SN~2014C are desirable in order to determine the epoch of disappearance of \Ha\ that will constrain the extent of the torus and the future evolution of the radio and X-ray emission. The X-ray flux is declining, suggesting that the main interaction of the shock with the CSM is over, in analogy with the behavior of SN~1987A. SN~2014C seems to be a more rapidly-evolving version of SN~1987A and hence may yield clues to the future behavior of SN~1987A.

Further observations are also encouraged to determine whether we are observing the effects of a pulsar wind nebula \citep{Milisavljevic12au}, as suggested by our observations of the \foiii velocity width and high excitation emission lines of [\ion{Fe}{7}] and [\ion{Fe}{10}].

The toroidal aspect of our interpretation is an integral concept of this paper and may apply to supernova and stellar evolution science far beyond the scope of SN~2014C.

\section*{Acknowledgements}

We thank the anonymous referee for a very thorough report that both clarified the paper and engendered some qualitatively new insights. We thank Kaew Tinyanont for sharing his NIR data and discussing issues of emission line profiles. We are grateful for support by the staff of McDonald Observatory and the Hobby-Eberly Telescope. 

BPT and JCW are supported in part by NSF grant 1813825, by a DOE grant to the Wooten Center for Astrophysical Plasma Properties (WCAPP; PI Don Winget), and by grant G09-20065C from the Chandra Observatory.
JV is supported by the project “Transient Astrophysical Objects” GINOP 2.3.2-15-2016-00033 of the National Research, Development and Innovation Office (NKFIH), Hungary, funded by the European Union.
VVD is supported by National Science Foundation grant 1911061 awarded to the University of Chicago (PI: Vikram Dwarkadas).
DP is supported in part by the National Aeronautics and Space Administration through Chandra Award Numbers GO0-11007A and GO GO9-20065A issued by the Chandra X-ray Center, which is operated by the Smithsonian Astrophysical Observatory for and on behalf of the National Aeronautics Space Administration under contract NAS8-03060.

The University of Texas at Austin sits on indigenous land. The Tonkawa lived in central Texas and the Comanche and Apache moved through this area. The Davis Mountains that host McDonald Observatory were originally husbanded by Lipan Apache, Warm Springs Apache, Mescalero Apache, Comanche and various tribes of the Jumanos. We acknowledge and pay our respects to all the Indigenous Peoples and communities who are or have been  a part of these lands and territories in Texas. 
 
\facilities{This study is based in part on observations made with the DIAFI camera mounted on the 2.7 m Harlan J. Smith telescope at McDonald Observatory. This study also employs observations obtained with the Hobby-Eberly Telescope, which is a joint project of the University of Texas at Austin, the Pennsylvania State University, Ludwig-Maximilians-Universit{\"a}t M{\"u}nchen, and Georg-August-Universit{\"a}t G{\"o}ttingen.
The HET is named in honor of its principal benefactors, William P. Hobby and Robert E. Eberly. The Low Resolution Spectrograph 2 (LRS2) was developed and funded by the University of Texas at Austin McDonald Observatory and Department of Astronomy and by Pennsylvania State University. We thank the Leibniz-Institut f{\"u}r Astrophysik Potsdam (AIP) and the Institut f{\"u}r Astrophysik G{\"o}ttingen (IAG) for their contributions to the construction of the integral field units. This study also utilized X-ray data from the {\it Neil Gehrels Swift Observatory}, \chandra, and \nustar\ and radio data from the {\it Karl G. Jansky Very Large Array}. The National Radio Astronomy Observatory is a facility of the National Science Foundation operated under cooperative agreement by Associated Universities, Inc.}

\software{ This research made use of; {\tt Astropy},\footnote{http://www.astropy.org} a community-developed core Python package for Astronomy \citep{astropy:2013, astropy:2018}; {\tt emcee} \footnote{https://emcee.readthedocs.io/en/stable/}, an MIT licensed pure-Python implementation of Goodman \& Weare’s Affine Invariant Markov chain Monte Carlo (MCMC) Ensemble sampler \citep{emcee:2013}; and the {\tt numpy}, {\tt scipy}, {\tt matplotlib} and {\tt pandas} python packages.
\swift\ data were reduced with {\tt XRTDAS} (v0.13.5), {\tt CALDB} (v20190910), {\tt XRTPIPELINE} and {\tt XSELECT}. \chandra\ data were processed with {\tt SPECEXTRACT}. \nustar\ data were processed with {\tt NUSTARDAS} (v20190812), {\tt CALDB}, and {\tt NUPIPELINE}. Spectral fitting was done with {\tt XSPEC} (v12.10.1f). VLA data were processed by the NRAO Pipeline for VLA observations using {\tt CASA}. }

\bibliographystyle{mnras}

\bibliography{main}

\appendix
\counterwithin{figure}{section}
\counterwithin{table}{section}

\section{Posterior distribution of the multi-component Gaussian model}
\label{sec:posterior}

In \S \ref{subsubsec:profiles} we derived multi-component Gaussian fits to the \Ha\ and other emission line profiles. We used the python package {\tt emcee} to perform a full MCMC fit and derive the relevant posterior distributions for each parameter. In the case of the \Ha\ profile, we used four Gaussians (with the exception of some of the earlier public data, for which we used five). There are thirteen parameters to the majority of the \Ha\ fits: the centroid $\mu$, the standard deviation $\sigma$, and the amplitude {\it A} of each Gaussian, as well as an overall baseline parameter {\it D}. 

\begin{figure}
    \centering
    \includegraphics[width=\textwidth]{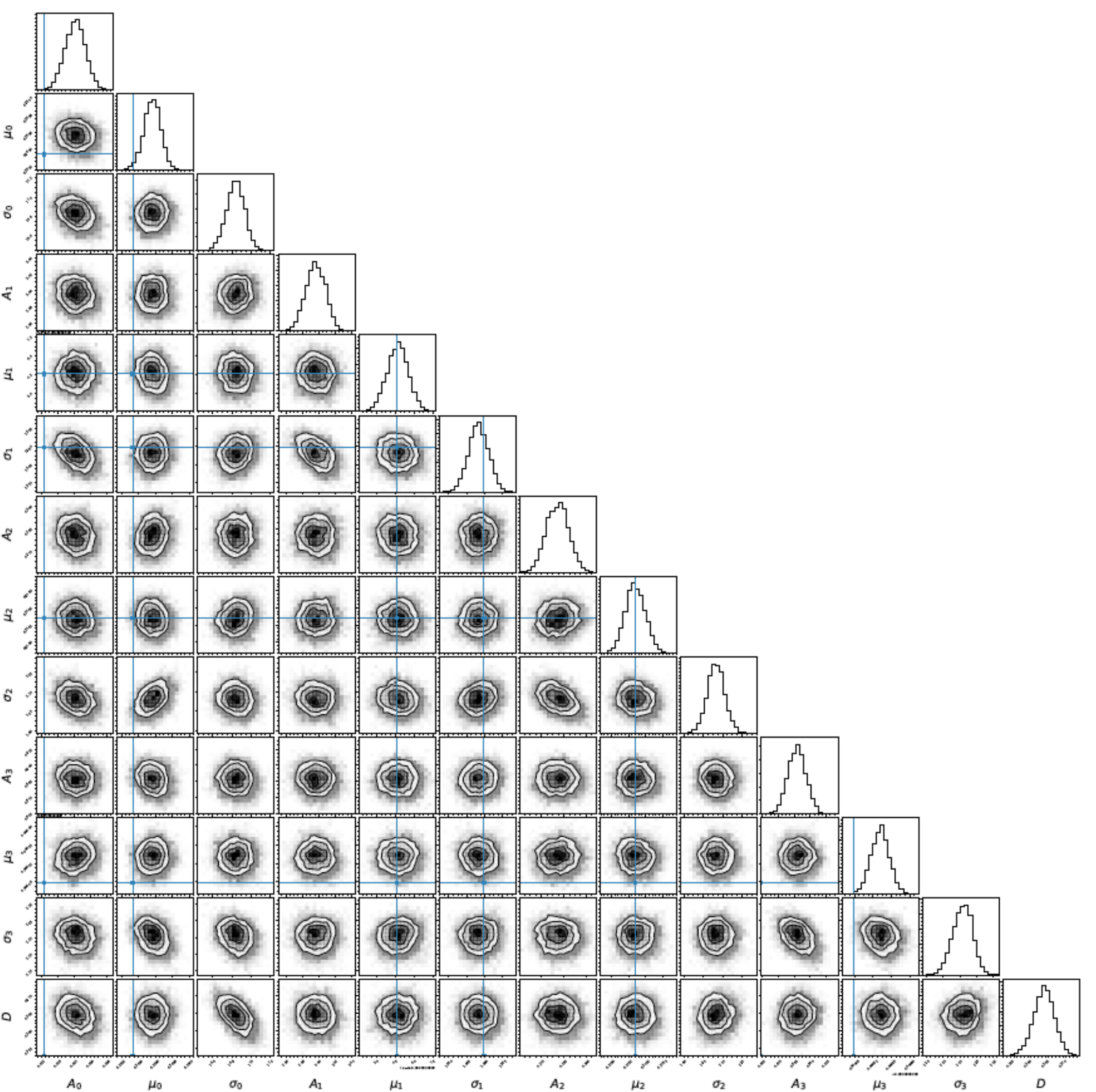}
    \caption{The full posterior distribution of our multi-component Gaussian model fits to the \Ha\ profile at $\phi = 1322$~d. The marginalized posterior probability distributions are shown across all pairwise matchings of fit parameters. The one-dimensional marginalized posteriors are shown on the top diagonal.   Parameter columns are in groups of three (triplets) representing the amplitude $A$, mean $\mu$, and standard deviation $\sigma$ of the individual Gaussian components. The first triplet of columns are $A$, $\mu$, and $\sigma$ for the broadened \Ha\ component, the second triplet are the same parameters but for the narrow \Ha\ component, the third and fourth triplet are those fit parameters for the two \fnii lines. The final column represents the baseline parameter that accounts for extraneous continuum flux. Blue lines indicate the initialization position obtained with a simple least-squares analysis. The order of parameters on the vertical axis (rows from top to bottom) is identical to the order on the horizontal axis (columns from left to right) described above.}
    \label{fig:posterior}
\end{figure}

We used thirty MCMC walkers for 500 steps including a burn-in phase of 300 steps. We used uniform priors for each parameter with reasonable ranges: 0.1 - 10 times the initial guess that was set by visually inspecting the data. An example posterior distribution is shown for the \Ha\ emission line at $\phi = 1322$~d is shown in Figure \ref{fig:posterior}. We use these posteriors to derive our estimate of quantities such as the luminosity and FWHM velocities and their associated statistical error from the fit. We find that these fit errors are subdominant relative to other sources of systematic error, such as the flux normalisation from the spectral calibration for the luminosities and the spectral resolution for the FWHM velocity widths.

\end{document}